%% file: h2galcosmoV2-01.tex
\documentclass[useAMS,usenatbib]{mn2e}

\usepackage{natbib}
\usepackage{aas_macros}
\usepackage{float}
\usepackage{amssymb}
\usepackage{amsmath}
\usepackage{tensor}
\usepackage{url}
\usepackage{wasysym}
\usepackage{graphicx}
\usepackage{epsfig} 
\usepackage{threeparttable} 
\usepackage{threeparttablex}
\usepackage{fancyhdr} 
\usepackage{longtable}
\usepackage{environ}
\usepackage{nomencl}
\usepackage{etoolbox}
\usepackage{lscape}
\usepackage{scrtime}

\usepackage[caption = false]{subfig}

\newcommand{\ha}{\mathrm{H}\alpha}
\newcommand{\hb}{\mathrm{H}\beta}

\newcommand{\Oiii}{\mathrm{O\ III}}

\newcommand{\hii}{H\,\textsc{ii}}

\newcommand{\mincir}{\raise-3.truept\hbox{\rlap{\hbox{$\sim$}}\raise4.truept\hbox{$<$}\ }}
\def\mean#1{\left< #1 \right>}

\newcommand{\ergs}{erg s$^{-1}$}                            

\title[Cosmology with high redshift HII galaxies]{On the road to precision cosmology  with high redshift HII galaxies\thanks{ Partly based on observations obtained at the European Southern Observatory,  programme ID 091.A-0413(A)} 
}
\author[R. Terlevich et al.]
{ R.~Terlevich,$^{1,2}$\thanks{E-mail: rjt@inaoep.mx \& rjt@ast.cam.ac.uk} 
   E.~Terlevich,$^{1}$  J.~Melnick,$^{3,8}$ R.~Ch{\'a}vez,$^{1}$ M.~Plionis,$^{4,1,5}$ F.~Bresolin,$^{6}$ 
   \newauthor and S.~Basilakos.$^{7}$ \\ \\
$^{1}$Instituto Nacional de Astrof{\'i}sica {\'O}ptica y Electr{\'o}nica, AP 51 y 216, 72000, Puebla, M{\'e}xico.\\
$^{2}$Institute of Astronomy, University of Cambridge, Madingley Road, Cambridge CB3 0HA, UK.\\
$^{3}$European Southern Observatory, Alonso de Cordova 3107, Santiago, Chile.\\
$^{4}$Physics Dept., Aristotle Univ. of Thessaloniki, Thessaloniki 54124, Greece \\
$^{5}$National Observatory of Athens, P. Pendeli, Athens, Greece \\
$^{6}$Institute for Astronomy of the University of Hawaii, 2680 Woodlawn Drive, 96822 Honolulu,HI USA.\\
$^{7}$Academy of Athens, Research center for Astronomy and Applied Mathematics, Soranou Efesiou 4, 11527, Athens, Greece.\\
$^{8}$Observatorio Nacional, Rua Jos\'e Cristino 77, 20921-400 Rio de Janeiro, Brasil. 
}

\begin{document}

\date{v2.01 ---- Compiled at \thistime\ hrs  on \today\  }

\pagerange{\pageref{firstpage}--\pageref{lastpage}} \pubyear{2014}

\maketitle

\label{firstpage}

\begin{abstract}

We report the first results of a programme aimed at studying the  properties of high redshift galaxies with on-going massive and dominant episodes of star formation (HII galaxies). We use  the $L(\hb) - \sigma$ distance estimator based on the correlation between the ionized gas velocity dispersions and Balmer emission line luminosities of HII galaxies and Giant HII regions to trace the expansion of the Universe up to $z \sim 2.33$. This approach provides an independent constraint on the equation of state of dark energy and its possible evolution with look-back time.
 
 Here we present high-dispersion (8,000 to 10,000 resolution) spectroscopy of HII galaxies at redshifts between 0.6 and 2.33, obtained at the VLT using XShooter. Using six of these HII galaxies we obtain broad constraints on the plane $\Omega_m - w_0$. The addition of 19 high-z HII galaxies from the literature improves the constraints and highlights the need for high quality emission line profiles, fluxes and reddening corrections. The 25 high-z HII galaxies plus our local compilation of 107 \hii\ galaxies up to $z=0.16$ were used to impose further constraints.  Our results are consistent with recent studies, although  weaker due to the as yet small sample and low quality of the literature data of high-z HII galaxies.
 
 We show that much better and competitive constraints can be obtained using a larger sample of high redshift HII galaxies with high quality data that can be easily obtained with present facilities like KMOS at the VLT.



\end{abstract}


\section{Introduction}

\hii\ galaxies  (HIIG) are  most extreme narrow emission line star forming systems selected  from spectroscopic surveys as those having the largest equivalent widths (EW) in their emission lines, i.e. EW(H$\beta) > $ 50\AA\  (or EW(H$\alpha) > $ 200\AA)  in their rest frame  and being extremely compact.
The  lower limit required in the selected equivalent width of the recombination hydrogen lines  is of fundamental importance to guarantee that the sample of HIIG is composed by systems in which a single and very young starburst,  less than 5 Myr in age,  dominates the total luminosity output. This selection criterion also minimizes the possible contamination by an underlying older population or older clusters inside the spectrograph aperture \citep[cf.][]{Melnick2000, Dottori1981, Dottori1981b, Chavez2014} and the ionizing photon escape. HIIG
 thus selected are  spectroscopically  indistinguishable from young Giant Extragalactic \hii\ Regions (GEHRs) in nearby galaxies (e.g. 30 Doradus in the LMC or NGC~604 in M33). This is highlighted by the fact that when discovered, the prototypical HIIG I~Zw18 and II~Zw40 were named  ``Isolated Extragalactic \hii\ regions" \citep{Sargent1970}.  On the other hand, the compactness requirement biases  the sample towards single bursts with relatively small crossing times.

Under the working hypothesis of instantaneous coeval star formation (ISF) it is possible to define an age for the stellar population. 
Even if  continuos star formation (CSF) is assumed and a wide range of stellar ages is present  the ionization will be dominated by the stars born in the last 5 Myrs. Thus from this point of view not much difference is expected in the observables of these extreme emission line starbursts between the instantaneous 
and the continuos case. There are parameters however that will differ like the EW of the IR-CaII triplet or the total L(H$\alpha$/Mass relation.
Another important difference is that in the case of CSF one expects to detect the WR features in all systems while for ISF they are expected only in a fraction of them.

 The optical properties of HIIG can be considered as those of a ``naked" extremely young and compact burst of star formation rather than those of the host galaxy.

It has been shown that GEHR  and HIIG exhibit a tight correlation between the luminosity and the width of their emission lines, the $L(\mathrm{H}\beta)-\sigma$
relation \citep{Terlevich1981}. 
The scatter in this  relation is small enough that it can be used to determine cosmic distances independently of redshift 
\citep[see][]{Terlevich1981,Melnick1988,Melnick2000, Fuentes-Masip2000,Telles2003,Bosch2002,Siegel2005,Bordalo2011,Chavez2012,Chavez2014}. 
HIIG can reach $\hb $ luminosities larger than $10^{42}$ \ergs\  making them observable even at relatively  large redshifts ($z > 3$) with present-day NIR spectrographs.

In (CSF) we showed that  the $L(\mathrm{H}\beta)-\sigma$  relation constitutes a viable alternative to SNe~Ia for the determination of cosmological parameters and presented a general strategy to use high $z$ HIIG as effective cosmological probes to reduce significantly the parameter space of the dark energy equation of state and test its possible evolution with redshift.

To date, the cosmic acceleration has been traced directly only by means of SNe Ia and  up to  redshift $z \sim 1.5$ \citep{Riess1998, Perlmutter1999, Amanullah2010, Hicken2009, Suzuki2012}, a fact which implies that it is important to use alternative probes at higher redshifts in order to verify the SNe Ia results and to obtain more stringent constraints in the cosmological parameters solution space, with the ultimate  aim of discriminating among the various theoretical alternatives which attempt to explain the accelerated expansion of the Universe \citep[cf.][]{Suyu2012}.

 In this paper we demonstrate the feasibility of  using HIIG as competitive cosmological tracers to high-$z$ thanks to the availability in 8m class telescopes of highly efficient near infrared high spectral resolution spectrographs. 
 
 We would like to reinforce what we consider a central point. The fact that a $L(\mathrm{H}\beta)-\sigma$  relation exists at all, means that it  can be used, empirically,  as a distance estimator up to large redshifts, regardless of  the physics that causes it. This aspect   has been discussed in the literature   \citep[among others][]{Terlevich1981,Melnick1988,GTT1993,Terlevich1997, Melnick1999, Melnick2000,Zaragoza-Cardiel2015} and will be discussed still  further as more and better data becomes available.

  In \S 2 we present the observations and data reduction, and the data (both ours and from the literature) are analysed in \S 3. The results are discussed in \S 4 and conclusions and plans for future work are given in \S 5.
 
\section{Observations and Data Reduction}

High spectral resolution spectroscopic observations were obtained using the XShooter  spectrograph \citep{Vernet2011} at the Cassegrain focus of the ESO-VLT (European Southern Observatory Very Large Telescope) in Paranal, Chile during the nights of 29 and 30 September 2013\footnote{Observing programme ID 60.A-9022(C)}. We obtained spectra for the three arms using a 0.6 arcsec slit, the typical spectral resolution on the VIS arm was $\sim 10000$, whereas in the NIR arm it was $\sim 8000$. Total exposure times per object ranged between 1 and 3 hours.

A sample of 9 star forming galaxies with rest frame equivalent width (EW) of H$\beta$~ $> 50\ \mathrm{\AA}$ or EW(H$\alpha$)~$> 200\ \mathrm{\AA}$ was selected from \citet{Hoyos2005, Erb2006b, Erb2006} and  \citet{Matsuda2011}. The redshift range covered was $0.64 \leq z \leq 2.33$. 
The observation parameters are detailed in  Table \ref{tab:tab00}. 


A narrow slit was needed to achieve the required resolution but when combined with the excellent seeing resulted in pointing problems associated with the  fact that these are faint and almost stellar objects. All pointings were blind and we had to rely on our differential astrometry. This proved to be correct, i.e.~inside $0.2$ arcsec in all cases but one.  The narrow slit has another negative effect in the sense that slight offsets or seeing becoming worse than the slit size result in light losses.
These  aspects highlight the importance of  spectroscopy with integral field units that minimizes the astrometry problems while at the same time optimizing the light collection.

The data reduction was carried out using the XShooter pipeline V2.3.0 over the GASGANO V2.4.3 environment\footnote{GASGANO is a JAVA based Data File Organizer developed and maintained by ESO.} using the `physical model mode' reduction.

\section{Data Analysis}
\subsection{Emission line widths}

\begin{figure*}
\centering

\subfloat{\label{test:1}\includegraphics[width=92mm]{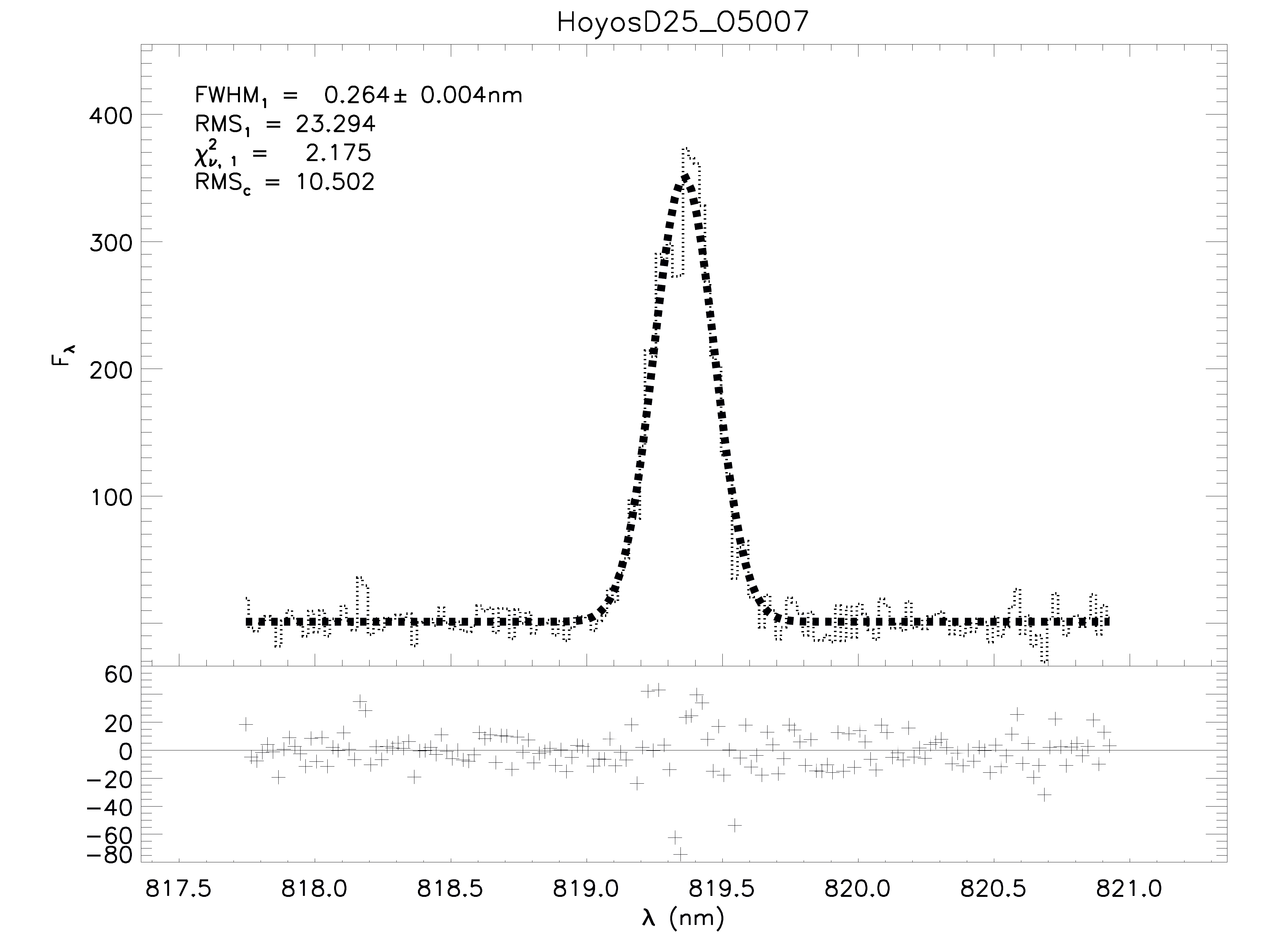}}
\subfloat{\label{test:2}\includegraphics[width=92mm]{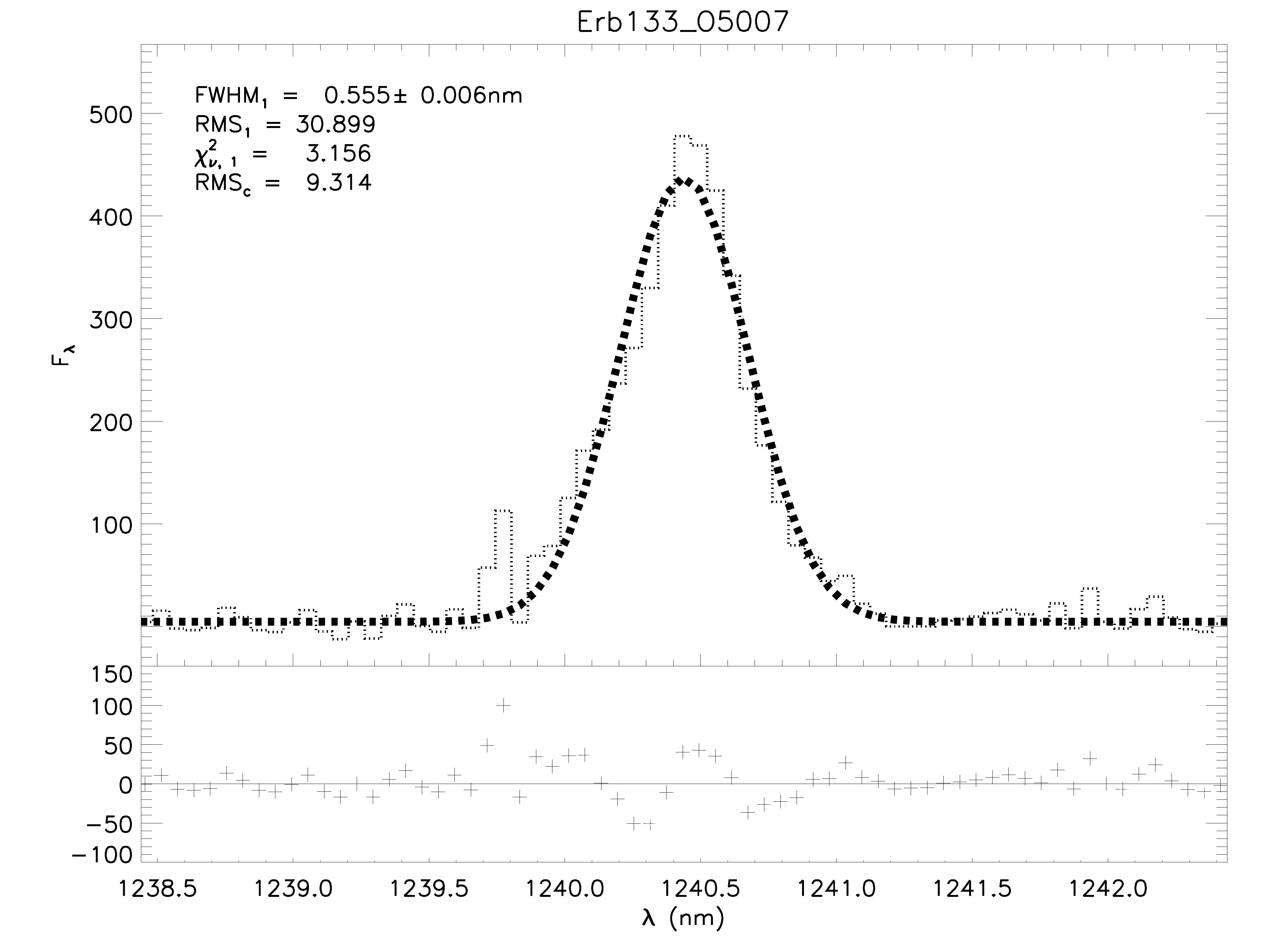}}\\
\subfloat{\label{test:3}\includegraphics[width=92mm]{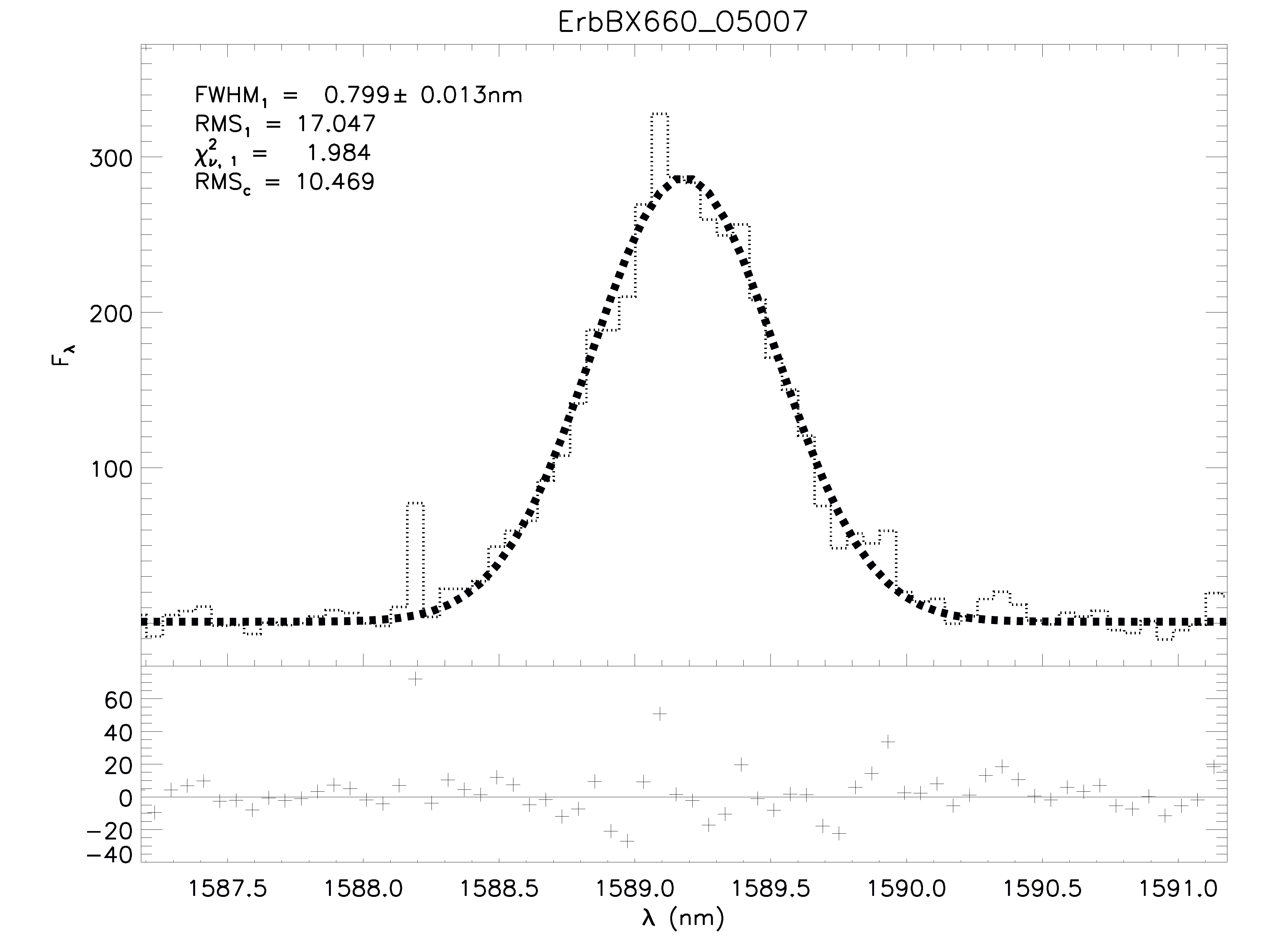}}
\subfloat{\label{test:4}\includegraphics[width=92mm]{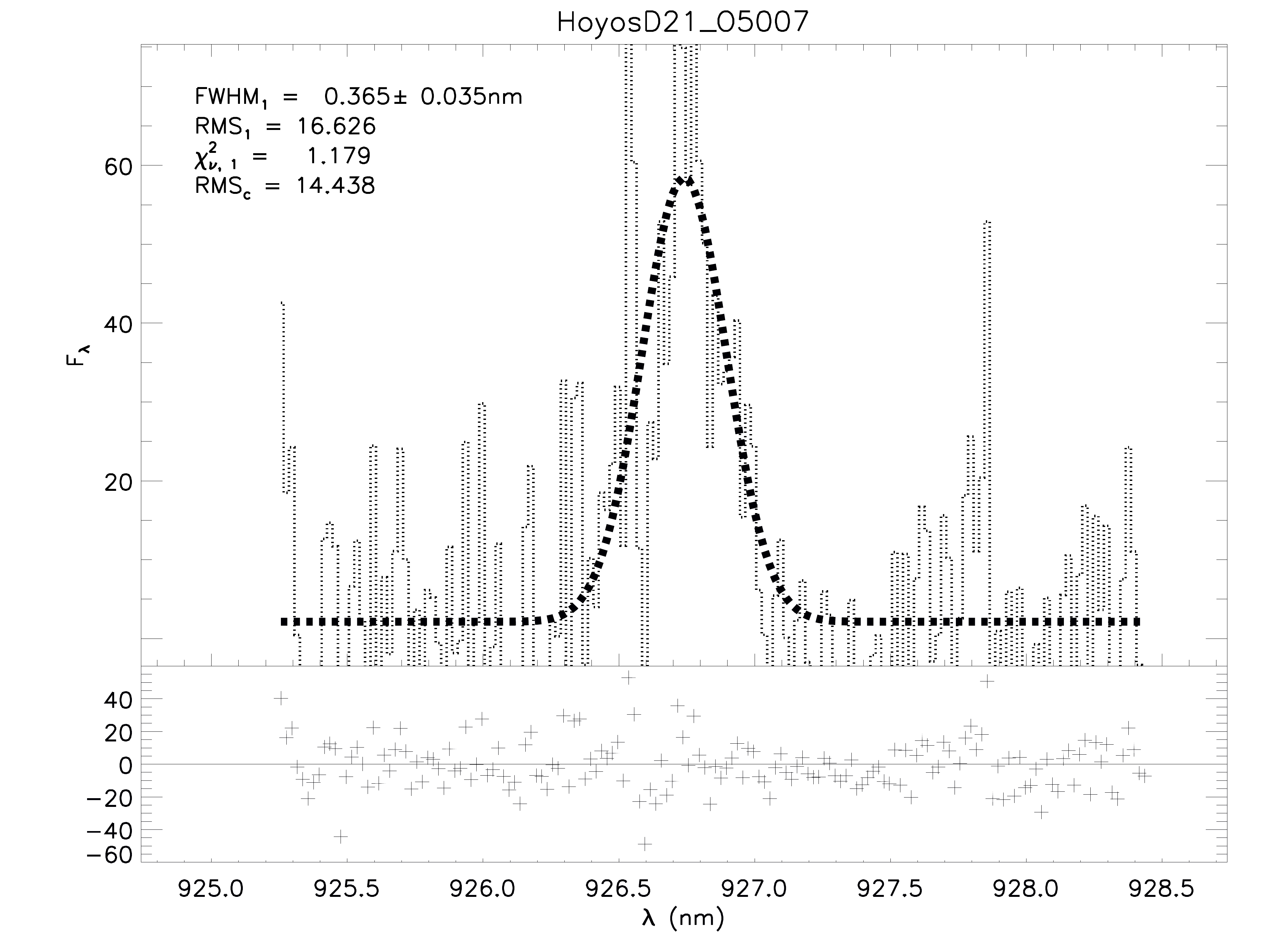}}\\
\subfloat{\label{test:5}\includegraphics[width=92mm]{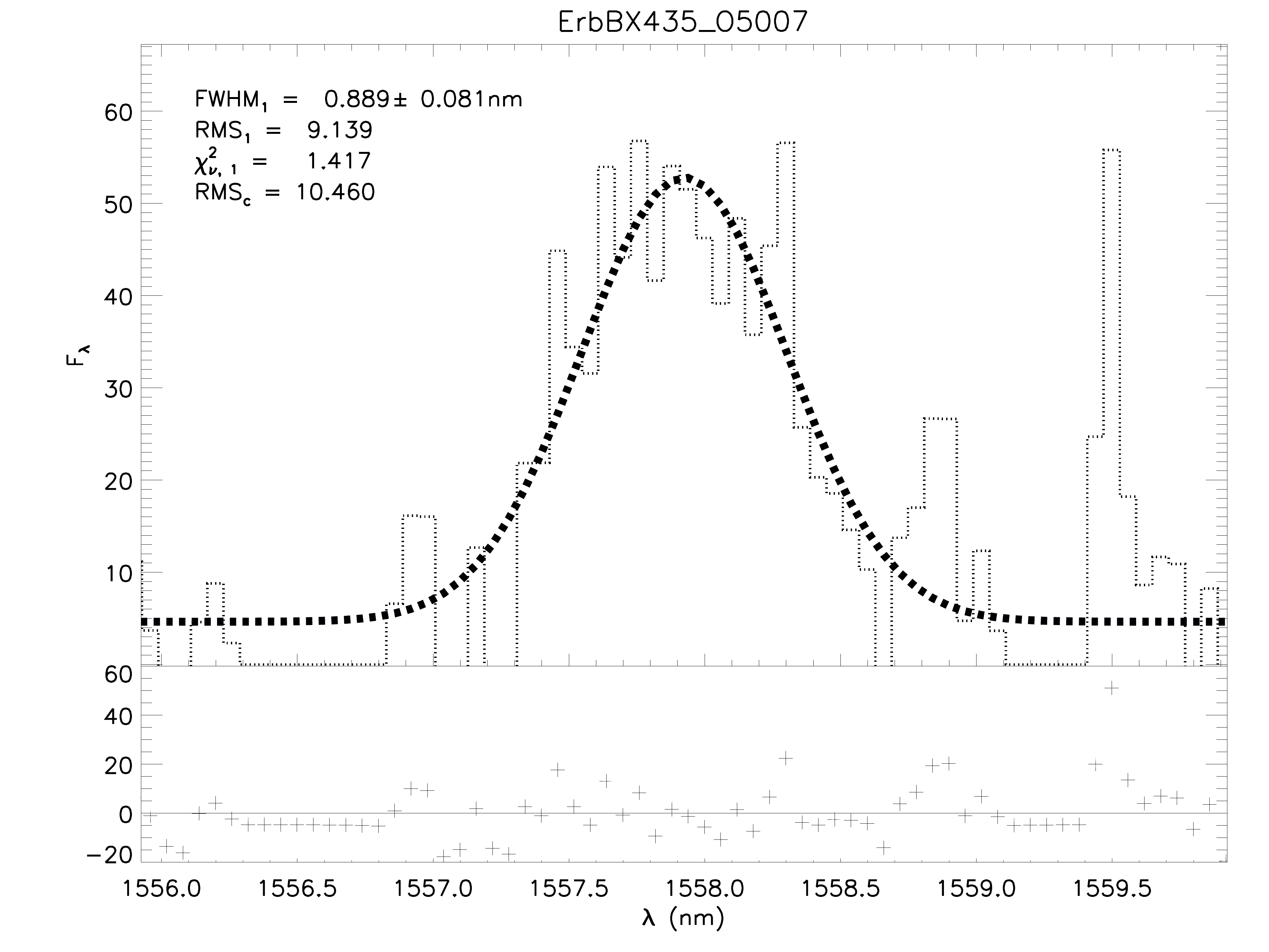}}
\subfloat{\label{test:6}\includegraphics[width=92mm]{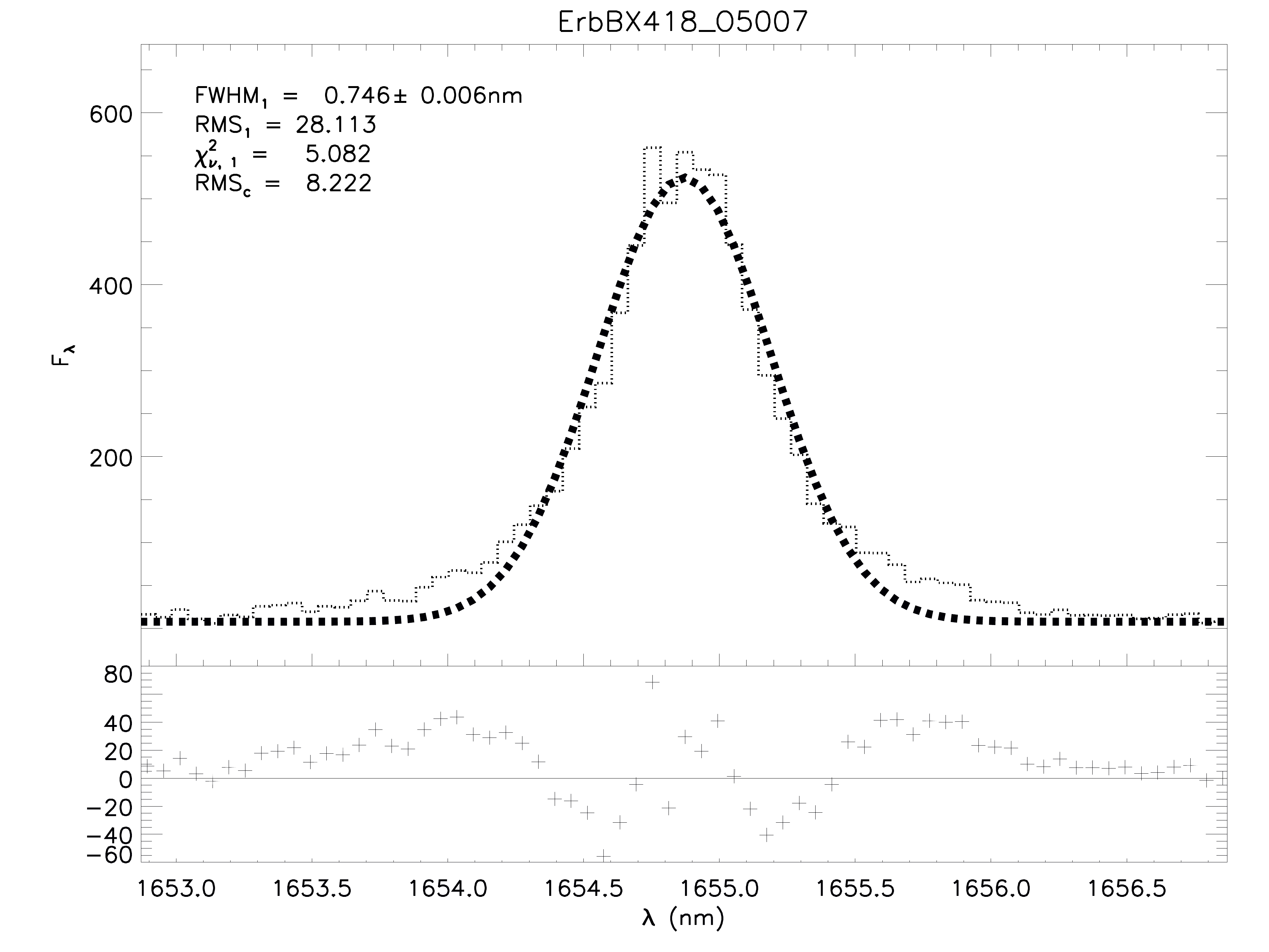}}\\

\label{fig:Gauss}
\end{figure*}

To determine the FWHM of the emission lines, single gaussians were fitted to the 1D spectral profiles of the $[\Oiii]\ \lambda 5007$\AA\  and $\ha$ lines when available\footnote{Historically, $\hb$ has been used but, when available, the stronger $\ha$ line is preferred}. These fits were performed using the IDL routine \verb|gaussfit|. Figure \ref{fig:Gauss} shows the fits  to the $[\Oiii]\ \lambda 5007$\AA\ line.

\begin{figure*}
\centering

\subfloat{\label{test:7}\includegraphics[width=92mm]{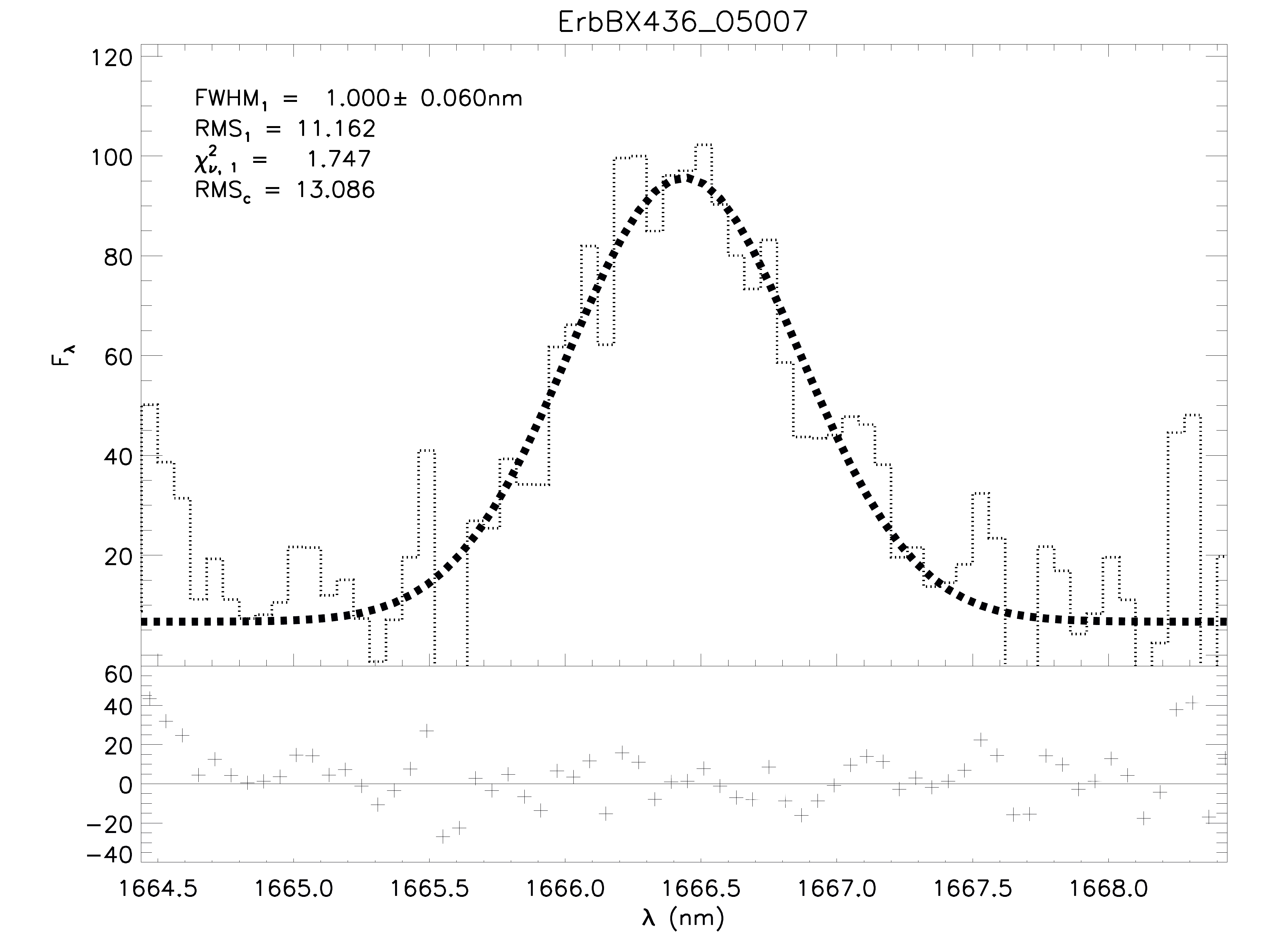}}
\subfloat{\label{test:8}\includegraphics[width=92mm]{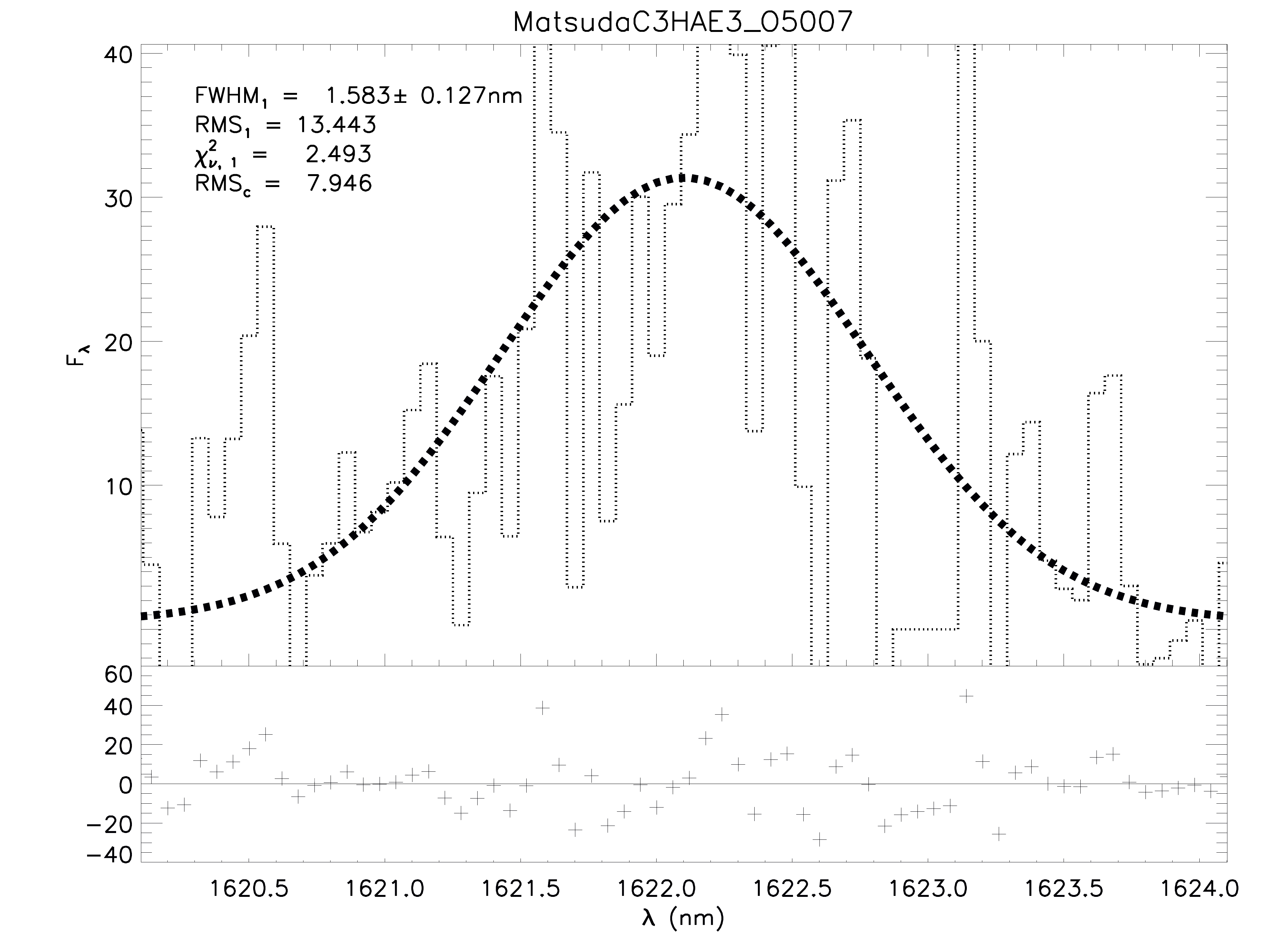}}\\
\subfloat{\label{test:9}\includegraphics[width=92mm]{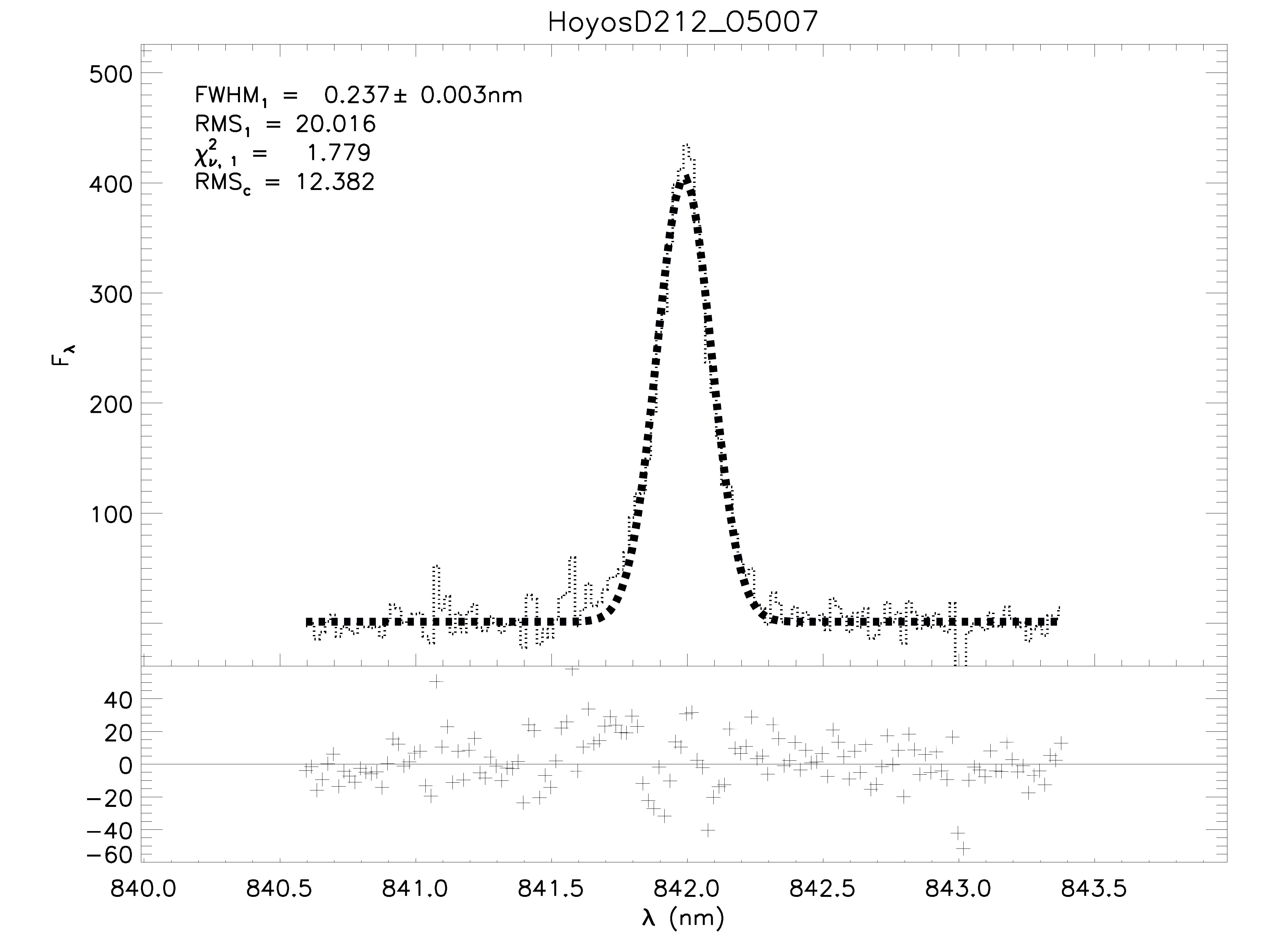}}


\caption{Gaussian fits to the $[\Oiii]\ \lambda 5007$\AA\ line for the  nine observed objects as labelled. \emph{Upper
    panel:} The single gaussian fit is shown with a  dashed line (thick black).
  The parameters of each fit are shown in the top left corner. \emph{Lower
    panel:} Residuals from the fit.}
\label{fig:Gaussb}
\end{figure*}

The uncertainties of the measured FWHM were estimated using a Montecarlo 
analysis. A set of 
random realizations of every spectrum was generated 
using the data poissonian 1$\sigma$ 1-pixel uncertainty. Gaussian fittings for every 
synthetic spectrum in the set were performed afterwards, and we obtained a distribution of 
``synthetic" FWHM 
from which the 1$\sigma$ uncertainty for the widths measured in the 
spectra follows.

Table \ref{tab:tab601} lists the FWHM measurements for the high
resolution observations prior to any correction such as instrumental
or thermal broadening. Column (1)  is the target 
name, column (2) is the heliocentric redshift as measured from the observed emission lines, columns (3) to  (6) contain  
the measured central wavelength in $\mathrm{nm}$ and FWHM in $\mathrm{km\ s^{-1}}$ for $[\Oiii]\ \lambda 5007$ 
and $\ha$ respectively.

The observed velocity dispersions ($\sigma_o$) -- and their 1$\sigma$ uncertainties -- have  
been derived from  the FWHM measurements of the $[\Oiii]  \lambda 5007$\AA\  and $\ha$
lines  ($\sigma_o = 0.4247 \times  FWHM$).
Corrections for thermal ($\sigma_{th}$) and instrumental ($\sigma_{i}$) broadening have been applied, yielding a final velocity dispersion
\begin{equation}
 \sigma = \sqrt{\sigma_o^2 - \sigma_{th}^2 - \sigma_i^2 - \sigma_{fs}^2} 
\end{equation}
 with a spin broadening value, in the case of  $\ha$, of $\sigma_{fs}(\mathrm{H}\alpha)  =  2.4\ \mathrm{km\ s^{-1}}$   as detailed in 
\citet{Chavez2014}.


The mean relation between  the velocity dispersions of [OIII] and of H$\alpha$ [where we could measure both, including our own low-z HIIG sample from \citet{Chavez2014}] was found to be $\sigma (H\alpha)$ = $\sigma$([OIII]) + ($2.91 \pm 0.31)\ \mathrm{km\ s^{-1}}$. We used this expresion to estimate $\sigma$(H$\alpha$) for those objects with only $\sigma$([OIII]) measured.

As concluded by \citet {Melnick1988} and \citet{Chavez2014}, imposing an upper limit  to the velocity dispersion of $\log \sigma($H$\beta) \lesssim 1.8$ $\mathrm{km\ s^{-1}}$, 
minimizes the probability of including  rotationally  supported systems. Therefore from the observed sample we selected all  objects having $\log \sigma(\ha) < 1.8$ $\mathrm{km\ s^{-1}}$ thus reducing the sample to 6 objects as indicated in Table \ref{tab:tab601}, column (7). 

The adopted emission line velocity dispersions and their 1-$\sigma$ uncertainties are shown in Table \ref{tab:tab602}, column (5).

\input{tab/tab00}


\input{tab/tab01}
\input{tab/tab02}


\subsection{Fluxes}
For the objects selected from \citet{Erb2006b, Erb2006}, the H$\alpha$ fluxes were obtained from the literature directly and we can readily deduce from the reddening corrected $f(\mathrm{H}\alpha)$, the $f(\mathrm{H}\beta)$  from their theoretical ratio. The three objects from \citet{Hoyos2005} do not have a direct measurement of their line fluxes; Their line luminosities  were  obtained using  a rough estimate from their total blue luminosity and equivalent width \citep[see][]{Terlevich1981}:
\begin{equation}
	B_C = -2.5 \log [L_o(\mathrm{H}\beta) / EW_{\lambda}] + 79.7\ \;,
\end{equation}
where $B_C$ is the absolute blue continuum magnitude, $L_o(\mathrm{H}\beta)$ is the reddening corrected $\hb$ luminosity and $EW_{\lambda}$ is the  rest frame equivalent width of $\hb$. These three objects have a larger uncertainty in their estimated emission line luminosities.

When available the reddening (here taken as A$_V$) was obtained from the literature. The A$_V$ was derived from the publisehd  $E(B-V)$ using the value of R$_v =$ 4.05  given by  \citet{Calzetti2000}. For those objects where the reddening was not available the mean A$_V =$ 0.33 from our local sample, was adopted.
We have verified that the mean  values of A$_V$ for the local and high-z samples are compatible. 
 
 The top block of table \ref{tab:tab602}, column (7) shows the adopted $\hb$ fluxes and their 1$\sigma$ uncertainties, after internal extinction correction, the adoptedA$_V$ values are shown in column (8).

\subsection{Data from the literature}
To complement the data and compare results with a larger sample (albeit of lower quality velocity dispersions)  we have searched the literature for measurements of 
emission line FWHM of high-z  HIIG. Following strictly our selection criteria we were able to select a sample of 6 HIIG from \citet{Erb2006b, Erb2006}, 1 from \citet{Maseda2014} and 12 from \citet{Masters2014} for which $\sigma (\ha)$ and $f(\ha)$ are given, have line ratios and position in diagnostic diagrams corresponding to HII regions and EW(H$\beta) > $ 50\AA \ or EW(H$\alpha) > $ 200\AA\ plus Log$~\sigma (\ha) < 1.8 + 1$ sigma error. Only those objects with error less than 25\% in the measured velocity dispersion are included.
 
 The lower block of  Table \ref{tab:tab602} shows the data for the 19 objects selected from  the literature. 

\section{Discussion}
 
\subsection{The $L - \sigma$ relation}
Taking the concordance $\Lambda$CDM cosmology as the fiducial model we calculated the distances and hence the luminosities for all the objects in our sample (see subsection 4.3). The values of the luminosities are listed in Table \ref{tab:tab602}, column (6). 
As mentioned in section 3.1 three of the objects listed  in tables 1 and 2  are not included in the analysis because they do not fulfil the selection criterion on $\sigma$.

\begin{figure*}
	\includegraphics[width=175mm]{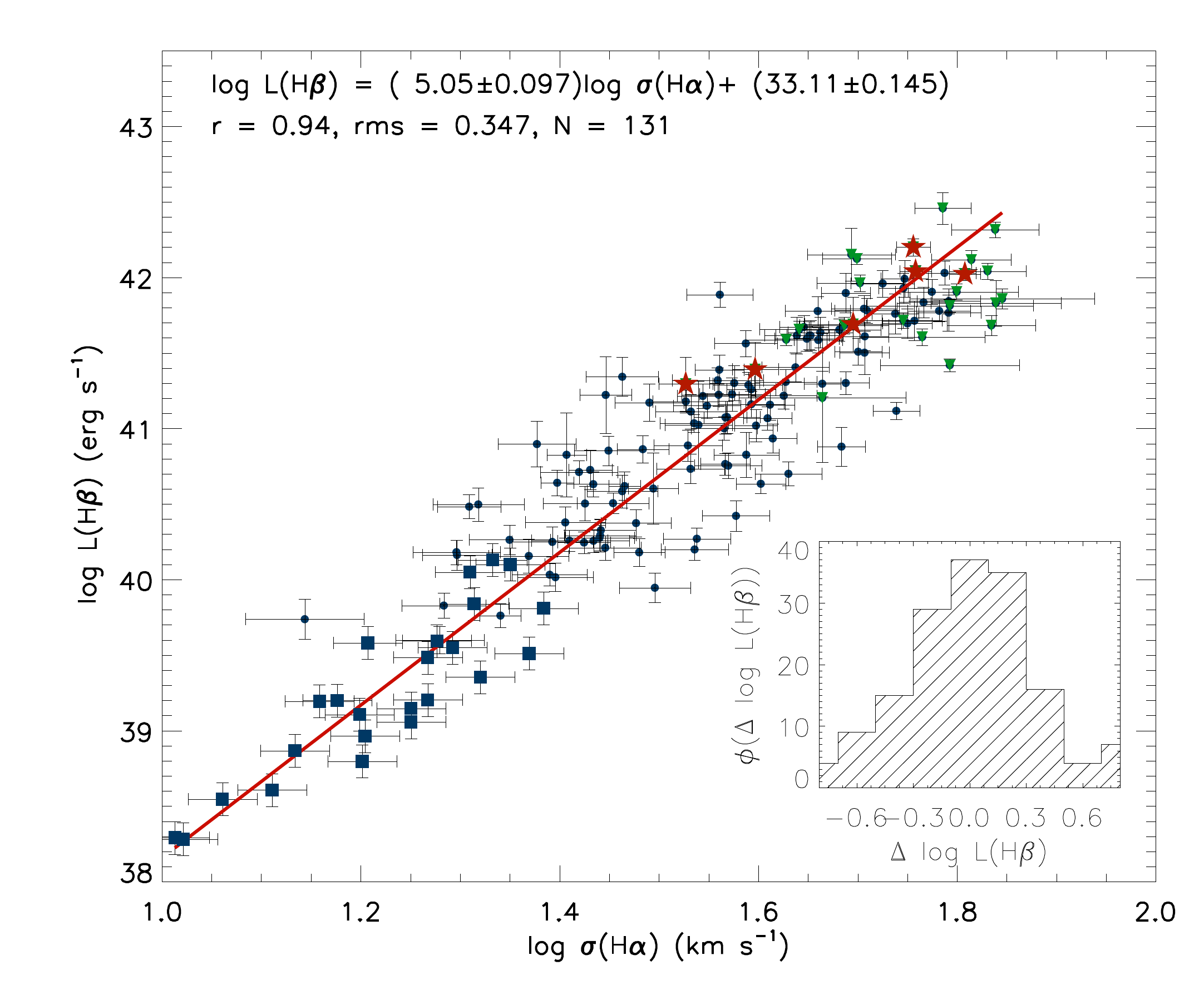}
 \caption{$L-\sigma$ relation for the combined local (131 HIIG and GHIIR) and high-z (25 HIIG) samples, the fit corresponds only to the local sample of 131 objects. Blue squares: GEHR. Blue dots:  local HIIG. Red stars: our high-$z$ XShooter observations. Green triangles: data from the literature. The inset shows the distribution of the residuals of the fit. The parameters of the fit are indicated at the top. }
\label{fig:LSTOT}
\end{figure*}

Figure ~\ref{fig:LSTOT} shows the $L - \sigma$ relation for the 25 high-z sample of \hii\ galaxies [6 high-z HIIG observed with XShooter (red stars) and 19  high-z HIIG from the literature (green triangles)],  and the local sample  of  GHIIR and  HIIG from \citet{Chavez2014}. 

{\bf The result is a remarkably tight   correlation  that underpins the use of the $L - \sigma$ relation as a distance estimator  over a wide range of distances, basically from the local group of galaxies (LMC, SMC, NGC~6822, M~33) up to at least z$\sim 2.3 $.}

Although here we are only considering the two dimensional $L - \sigma$ relation, we would like to point out  that according to \citet{Chavez2014} by including additional  observables in the $L - \sigma$ relation  like the size of the ionized gas region, the equivalent width of either $\hb$ or $\ha$ and the ionized gas metallicity or the continuum colour, the scatter  is substantially reduced from an rms$\sim$0.35 to an rms$<$0.25.
The importance of this reduction in the scatter of the distance estimator cannot be overemphasized.

\begin{figure*}
	\includegraphics[width=175mm]{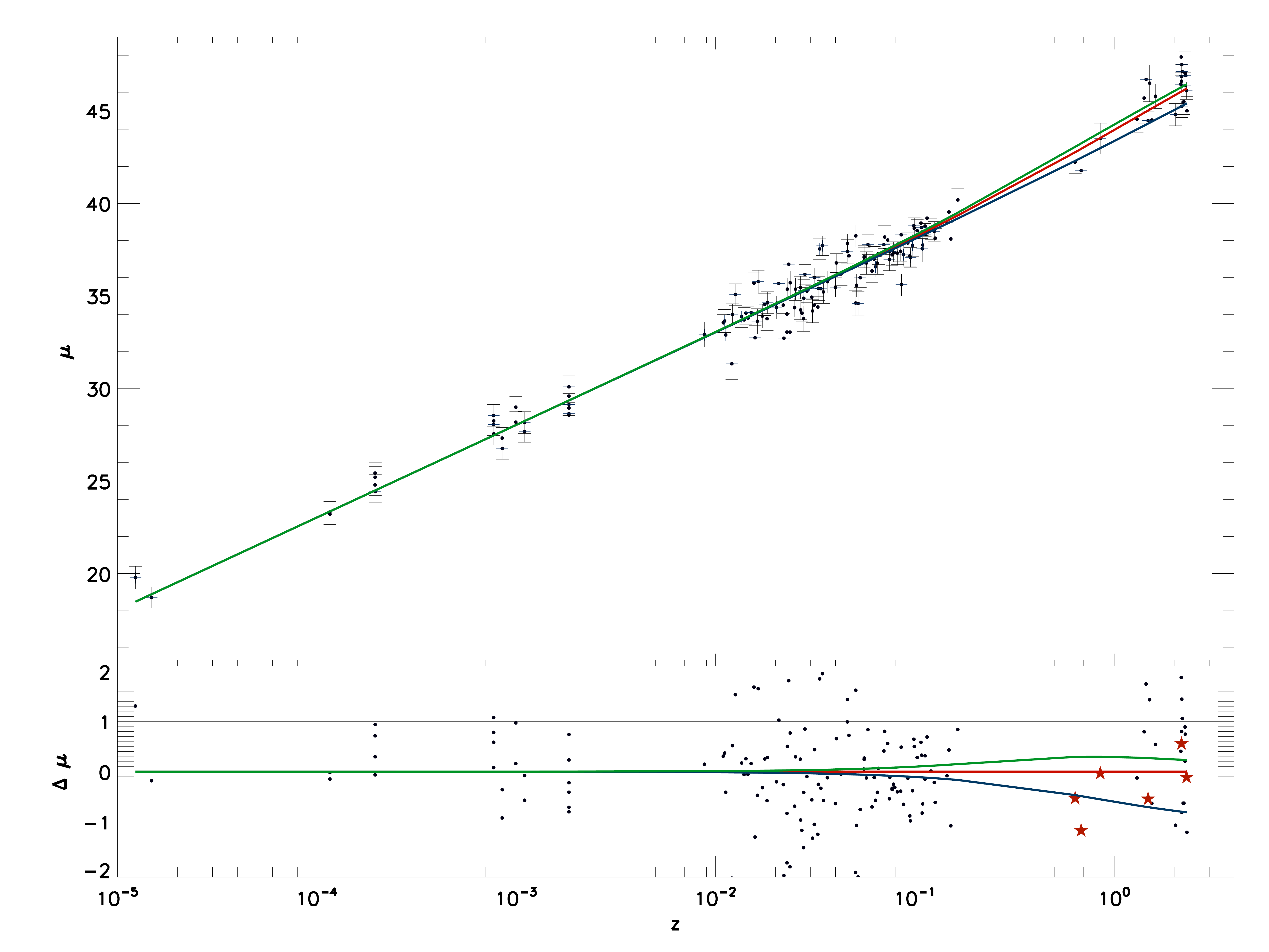}
 	\caption{Hubble diagram for our sample of low and high-z HIIG for three different cosmologies. The solid red line indicates the concordance $\Lambda$CDM cosmology with $\Omega_m = 0.3 ;  w_0\ = -1.0 $ and H$_0$=74.3. The solid green line shows a cosmology with $\Omega_m = 0.3$ and $w_0 = -2.0$. The solid blue line  corresponds to $\Omega_m = 1.0 $ and $\Omega_\lambda = 0.0$. In all three cases $ \Omega_k =0$. Residuals are plotted in the bottom panel. Note the huge dynamical range in distance modulus of almost 30 magnitudes covered with the $L - \sigma$ distance estimator.}
 	 \label{fig:HD}
\end{figure*}


\begin{figure*}
  \centering
  \subfloat[]{\label{test:2}\includegraphics[width=84mm]{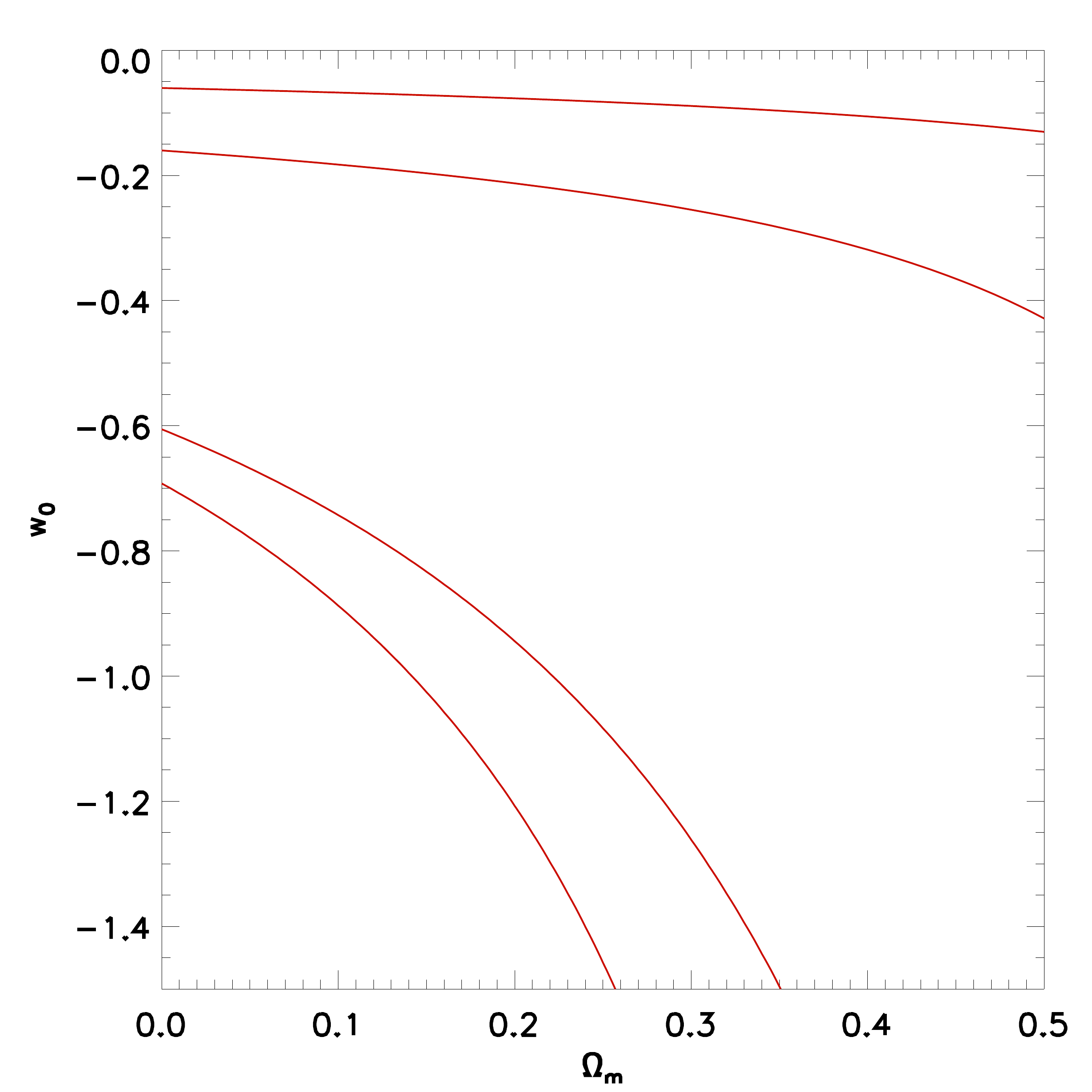}}
  \subfloat[]{\label{test:4}\includegraphics[width=84mm]{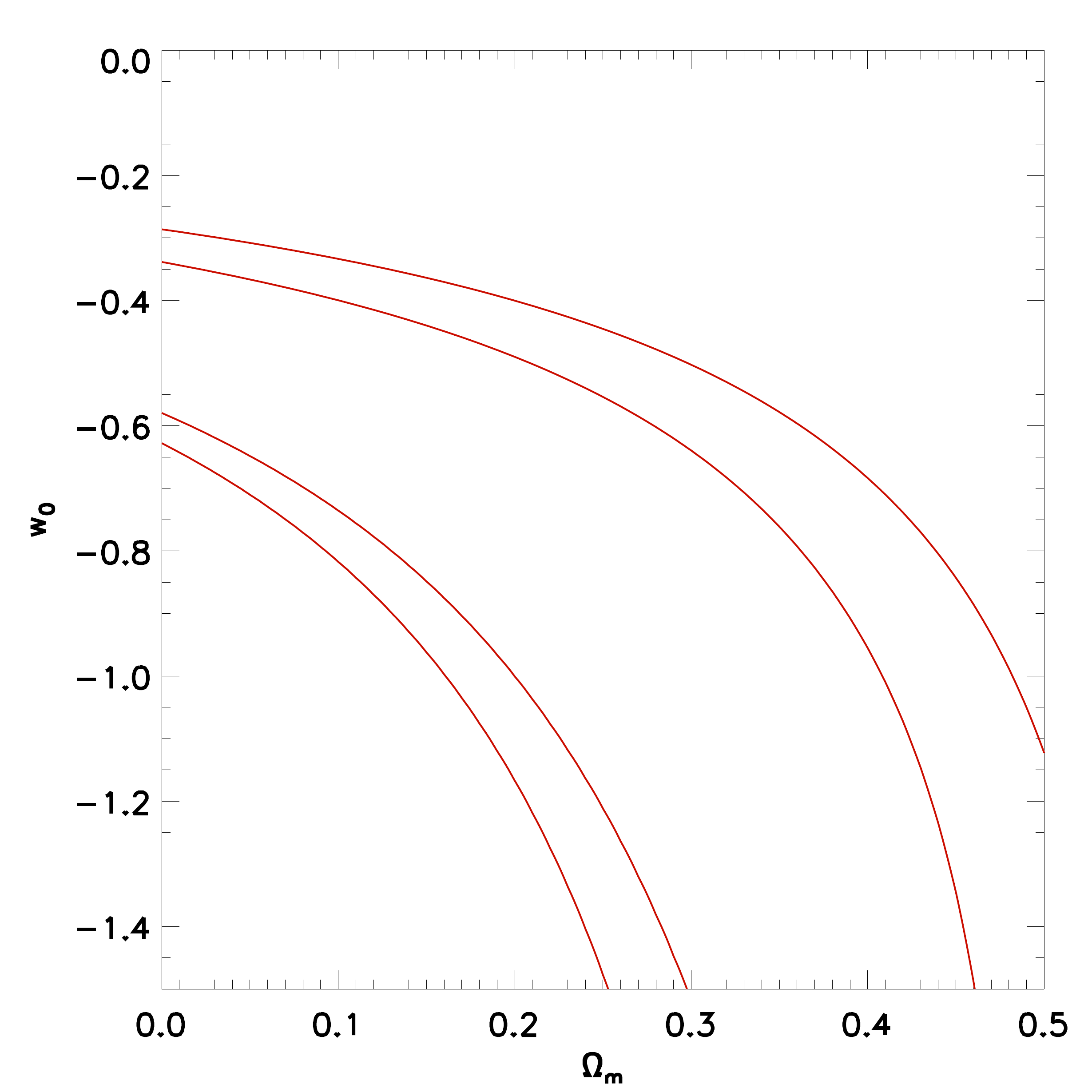}}
 \caption[Solution space  for $\mathbf{p} = \{\Omega_m, w_0\}$]{\small Solution space in the plane $ \{\Omega_m, w_0\}$ (see text). Panel (a)  for the 6 XShooter high-$z$ objects. Panel (b) Same as panel (a) including  19 high-$z$ objects  from the literature.  In both panels we show the 1 and 2$\sigma$ contours.}  \label{fig:Omw0}
\end{figure*}

\begin{figure*}
  \centering
  \subfloat[]{\label{test:1}\includegraphics[width=84mm]{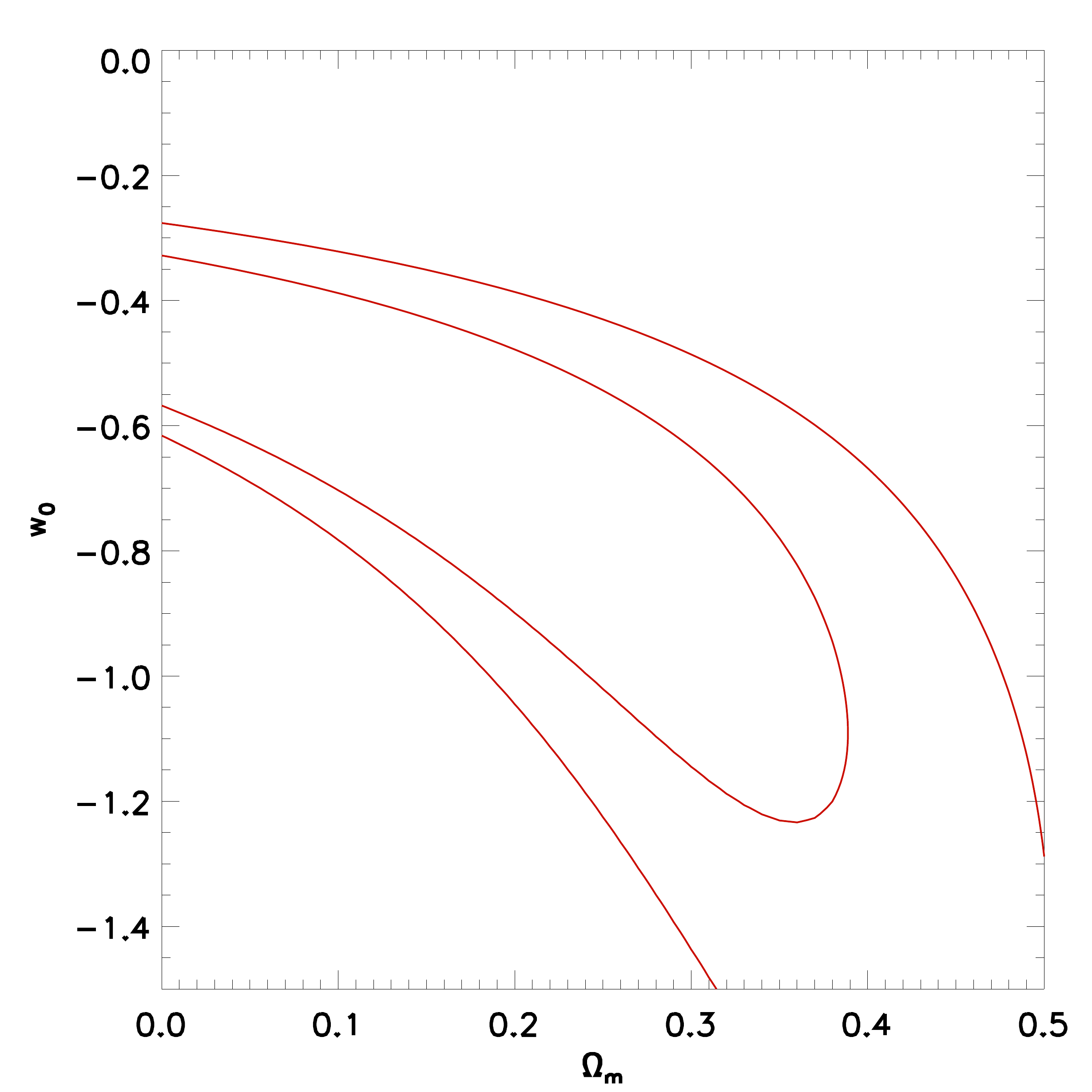}}
  \subfloat[]{\label{test:2}\includegraphics[width=84mm]{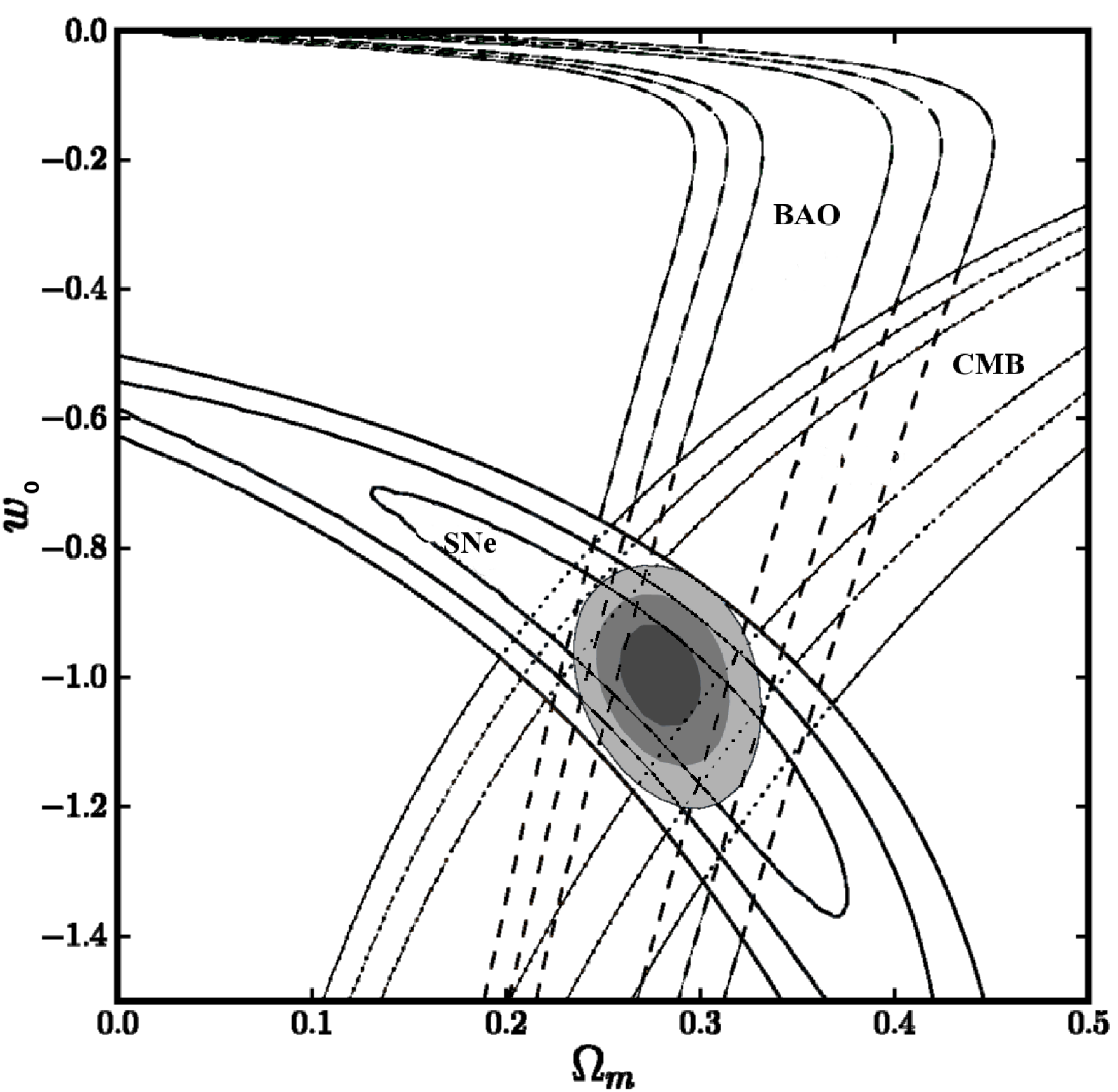}}
  \caption[Comparison on restrictions on the plane $\mathbf{p} = \{\Omega_m, w_0\}$]{\small Comparison of restrictions on the plane $\{\Omega_m, w_0\}$(see text). Panel (a) shows our results, obtained as described in the text for the combined 25 high-$z$ HIIG and the local  sample (131 HIIG and GHIIR). 1 and 2$\sigma$ contours (random) are shown. Panel (b) after \citet{Suzuki2012}
  shows the  recent results for 580 SNe Ia, CMB and BAOs, the 1, 2 and 3$\sigma$ contours  (random) are shown. }
  \label{fig:omwoco}
\end{figure*}


\subsection{The Hubble diagram}

Figure~\ref{fig:HD} shows the Hubble diagram for the joint sample of local and high-$z$ systems.
The points correspond to individual HIIG;  their distance moduli are obtained from:
\begin{equation}
	\mu^{ob} = 2.5 \log  L(\mathrm{H}\beta)_{\sigma} - 2.5 \log f(\mathrm{H}\beta) - 100.2
\label{eq:mu1}	
\end{equation}
where   $L(\mathrm{H}\beta)_{\sigma}$ is estimated from the $L-\sigma$ relation (Figure \ref{fig:LSTOT}) calculated for the joint local HIIG (107 objects) and GEHR (24 objects) samples, 
\begin{equation}
	\log L(\mathrm{H}\beta) = (5.05 \pm 0.097) \log \sigma (\mathrm{H}\alpha) + (33.11 \pm 0.145)
\label{eq:LS}	
\end{equation}

The continuous lines show the behviour of the theoretical distance modulus with redshift computed for three different cosmological models as:
\begin{equation}
	\mu^{th} = 5 \log D_L (\mathbf{p},  z_i) + 25, 
\label{eq:mu2}	
\end{equation}
where $D_L (\mathbf{p},  z_i)$ is the luminosity distance as calculated from a set of cosmological parameters, $\mathbf{p}$, and a given value of redshift, $z$.
The red line shows the  behaviour of the Concordance $\Lambda$CDM cosmology with H$_0$=74.3.  The green line shows the trend for a cosmology with $\Omega_m = 0.3$ and $w_0 = -2.0$. The solid blue line  corresponds to $\Omega_m = 1.0 $ and $\Omega_\lambda = 0.0$. In all three cases $ \Omega_k =0$ i.e Universe flatness is preserved.
The differential version of the Hubble diagram is shown in the bottom panel.

This is a remarkable and unique Hubble diagram in the sense that it covers a huge dynamical range with a single distance estimator. It connects galaxies in the local group  to galaxies at z$\sim$ 2.3, a range of almost 30 magnitudes in distance modulus or more than 5 dex in redshift.

\subsection {Towards precision cosmology with HII galaxies}

The most general set of cosmological parameters, assuming a flat Universe, a negligible value of the radiation density parameter $\Omega_r$ and using the Chevallier-Polarski-Linder (CPL) \citep{Chevallier2001, Linder2003} model for parametrizing the value of the dark energy equation of state parameter $w(z)$, is given by $\mathbf{p} = \{H_0, \Omega_m, w_0, w_1\}$ where $w_0, w_1\}$ are the first two terms of a Taylor expansion around the present epoch, namely
$w(a)=w_{0}+w_{1}(1-a)$ where $a=1/(1+z)$ is the scale factor of the Universe. In this case the luminosity distance used to estimate the value of $\mu^{th}_i(\mathbf{p},  z_i)$, is given by \citep[cf. e.g.][]{Weinberg2008, Frieman2003},
\begin{eqnarray}
	 D_L &=& c(1+z) \int_{0}^{z}{\frac{dz'}{H(z')}}\\
	 H^2(z,{\bf p}) &=& H_0^2 \big[ \Omega_m(1+z)^3 + (1 - \Omega_m) (1+z)^{3(1 + w_0 + w_1)}  \nonumber \\  
	 	     &&   \times \exp\left( \frac{-3 w_1 z}{z+1} \right) \bigg]\;.
\end{eqnarray}


To restrict the set of cosmological parameters  we  minimised the Likelihood function,
\begin{equation}
\chi^2(\mathbf{p}) = \sum\limits_{i=1}^n \frac{[\mu^{ob}_i(\sigma_i, f_i) -
  \mu^{th}_i(\mathbf{p},  z_i)]^2}{\sigma_{\mu^{ob}_i}^2 }, 
\end{equation}  
$\mu^{ob}_i(\sigma_i, f_i)$ are the `observed' distance moduli obtained from equation \ref{eq:mu1}; $\sigma_i$ are the measured velocity dispersions and  $f_i$ are the measured $\hb$ fluxes for each object.
$\mu^{th}_i(\mathbf{p},  z_i)$ are  the `theoretical' distance moduli from equation \ref{eq:mu2} obtained from
the measured redshifts  by using a particular set of cosmological parameters $\mathbf{p}$.
$\sigma_{\mu^{ob}_i}$ are their errors
propagated from the uncertainties in $\sigma_i$ and $f_i$ and the slope and intercept
of the distance estimator in equation \ref{eq:LS}. The summation is over the combined sample of  HIIGs,

Adopting the value of $H_0 = 74.3 \pm 3.1$ obtained in \citet{Chavez2012}, $w_1 = 0$  we obtain the results shown in Figure~\ref{fig:Omw0} for $\mathbf{p} = \{\Omega_m, w_0\}$ for the high-z  sample only.
Panel (a) shows the solution for  the 6 high-z HIIG observed with XShooter; 
Panel (b) shows the effect of including the 19 high-z objects from the literature. It is encouraging  that even for this small number of high-$z$ objects we are able to restrict so much the solution space. Although the literature data are of lower quality regarding the FWHM measurements, still by adding them we are  able to improve the result. This is a remarkable result considering the small number of intermediate to high-$z$ data (just 25 objects), of which only 6 have high quality ad-hoc observations.

In Figure~\ref{fig:omwoco}, we compare our results for the space  $\mathbf{p} = \{\Omega_m, w_0\}$, joining the  high-$z$ with the local  HIIG samples (see left panel), with recent results from SNe Ia, CMB and BAO (right panel). 
The figure shows the contrains of the properties of dark energy using SNe Ia alone \citep{Amanullah2010}, the seven-year Wilkinson Microwave Anisotropy Probe data of the CMB \citep{Komatsu2011}, the position of the BAO peak from the combined analysis of the SDSS Data Release 7 and 2dFGRS data \citep{Percival2010}. The combined restrictions from SNe Ia, CMB and BAO  and the measurement of the Hubble constant ($H_0$) from Cepheids \citep{Riess2011} are also shown.

It is clear from the figure that our present constraints on the space  $\mathbf{p} = \{\Omega_m, w_0\}$ are weaker than those for SNe Ia, but this is not surprising since in our case we have only 156 objects, most of them at z$<$0.16 a region of space where differences between Cosmological models are almost negligible,  vs.~580 SNe Ia with a maximum redshift of $\sim 1.5$.
 The strength of our results is that our sample includes 19 objects with $z > 1.5$ where the differences between models reach maximum values.
 
 From the comparison of the figures we can conclude first that there are no systematic shifts between the HII galaxies and SNe~Ia solutions and secondly that with a larger sample of HII galaxies with  high quality data it may be possible to achieve at least similar and probably even better results to those obtained with SNe~Ia as found in our simulations \citep{Plionis2011} and discussed below.


\subsection{The future of the $L(\hb) - \sigma$ distance estimator}

To estimate the number of high-$z$ tracers ($N_{Hz}$) required  to substantially reduce the cosmological parameters solution space we used the figure of merit (FoM) as in \citet{Plionis2011}. 
The FoM is the reciprocal area of the $2\sigma$ contour in the parameter space of any two degenerate cosmological parameters. In this way a larger FoM indicates better restrictions to the cosmological parameters. \citet{Plionis2011} use the parameter $S$ or ``reduction factor" to compare the ratio of FoM of SNe Ia  + high-$z$ tracers to that of only SNe Ia. They found that the number of high-$z$ tracers needed to obtain a given S in the quintessence dark energy (QDE) model implying that w is constant but different from -1,  
when combined with the intermediate and low-$z$ SNe Ia, can be expressed as:
\begin{equation}
	N_{Hz} \simeq 187 S/S_{100} - 88\, 
\end{equation} 
where $S_{100} = 1.87 \log (\mean{\sigma_{\mu}}^{-1} + 0.74) + 1.28$, and $\mean{\sigma_{\mu}}$ is the mean distance modulus error for the tracer. In the case of HIIG, ${\sigma_{\mu}} \simeq 0.6$,  $S_{100} \simeq 1.99$ and consequently in order to obtain a reduction of $S = 2$, we need around  100 high-$z$ HIIG. We can visualize this in Figure~\ref{fig:omwoco} panel (a) for the $\mathbf{p} = \{\Omega_m, w_0\}$ plane, where a factor of 2 reduction would mean that the future $2\sigma$ contours would be similar to the present $1\sigma$ ones. 

For  the CPL model where $w_1$ is variable  the same result  can be expressed as:
\begin{equation}
	N_{Hz} \simeq 404 S/S_{100} - 300\, 
\end{equation} 
where $S_{100} = 0.49 \log (\mean{\sigma_{\mu}}^{-1} + 0.65) + 1.09$. In the case of HIIG  $S_{100} \simeq 1.27$ and consequently in order to obtain a reduction of $S = 2$, about 300 high-$z$ HIIG would be required to achieve a result similar to that obtained with present day SN~Ia samples.

\section{Conclusions}

We have tested the use of high redshift  HIIG to trace the expansion of the Universe by means of their $L(\hb) - \sigma$ distance estimator.
To this end, we presented observations of a sample of only 9 HIIG in the redshift range of $0.6 \leq z \leq 2.3$ obtained with the ESO VLT XShooter spectrograph. After rejecting three HIIG due to either poor data or because they fall outside the selection window, we have used the remaining six  to obtain constraints on the $\{H_0, \Omega_m\}$ and  $\{\Omega_m, w_0\}$ planes. 

The results are surprisingly good considering the small number of objects and that only two nights of XShooter time were used. 
The addition of 19 HIIG from the literature (albeit of poorer quality) bringing the total number of objects to 25, did improve the results and provides a clear indication of how much an increase in number can improve the constraints on  the $\{H_0, \Omega_m\}$, $\{\Omega_m, w_0\}$ and$ \{w_0, w_1\}$ planes. 

Although our constraints are consistent with other determinations they are, as expected, definitely weaker. This is due to the small size of the intermediate to high-$z$ sample, and the considerable uncertainties in the data taken from the literature. 

Using the figure of merit approach we have estimated the expected improvement in the estimates of the cosmological parameters with a larger sample of high-$z$ HIIG. In particular, between 100 and 300 high-$z$ HIIG  are needed to obtain a reduction in the errors of at least a factor of two. 
A factor of 2 reduction implies that  the $2\sigma$ contours in Figure~\ref{fig:omwoco} panel (a) for the $\mathbf{p} = \{\Omega_m, w_0\}$ plane would be similar to the  present $1\sigma$ ones and  comparable to the $2\sigma$ SNe Ia contours in panel (b) that are the result of the analysis of 580 SNe Ia that took at least a decade and hundreds of nights of large telescope time to compile. In contrast, the observation of 100 HIIG can easily be achieved with multiple integrated field units instruments like VLT- KMOS,  in less than thirty hours observing given the relative abundance of HIIG with a considerable number  of them appearing in its 7.5 arcmin  field of view.

Finally the comparison of the results of SNe Ia and HIIG will undoubtedly contribute to learning about the systematic errors that limit the precision of both empirical methods while helping us to gain an insight into the intrinsic properties of HIIG at high redshifts. We also envisage that a substantial improvement to the present restrictions of cosmological parameters will be obtained by combining a few hundred HIIG with the  SNe Ia data. 

\section*{Acknowledgements}

We would like to thank the ESO time allocation committee for generously
awarding observing time for this project. We are indebted to Dawn Erb and Carlos Hoyos for providing all the help we needed for locating the high-z targets and are happy to acknowledge an anonymous referee whose comments helped to improve the clarity of this work. RT, ET, RC  and MP are grateful to
the Mexican research council (CONACYT) for supporting this research
under grants  CB-2005-01-49847, CB-2007-01-84746 and
CB-2008-103365-F  and studentship 224117.  
SB acknowledges support by the Research
Center for Astronomy of the Academy of Athens in the context of the
program {\it ``Tracing the Cosmic Acceleration''}. The hospitality of ESO (Chile), was gratefully enjoyed specially the help from the telescope support team, Julien 
Girard, Petr Kabath and Andr\'es Pino. 

\bibliography{bib/bibpaper2014}

\clearpage

\label{lastpage}

\end{document}

%% file: tab/tab00.tex
\begin{table*}
 \scriptsize 
 \centering
 \begin{minipage}{100mm}
  \caption{Observing log.}
  \label{tab:tab00}
  \begin{tabular}{@{} l c c c c c @{}}
\hline \hline \\[-2ex]
\multicolumn{1}{c}{(1)} &
\multicolumn{1}{c}{(2)} &
\multicolumn{1}{c}{(3)} &
\multicolumn{1}{c}{(4)} &
\multicolumn{1}{c}{(5)} &
\multicolumn{1}{c}{(6)} \\

\multicolumn{1}{c}{\bf{Name}} &
\multicolumn{1}{c}{\textbf{RA}} & 		
\multicolumn{1}{c}{ \textbf{$\mathbf{\delta}$}} & 		
\multicolumn{1}{c}{ \textbf{date}} & 	
\multicolumn{1}{c}{ \textbf{slitwidth}} & 	
\multicolumn{1}{c}{ \textbf{exp. time}} \\ 				

\multicolumn{1}{c}{} &
\multicolumn{2}{c}{(J2000.0)} &
\multicolumn{1}{c}{} & 
\multicolumn{1}{c}{(arcsec)} & 
\multicolumn{1}{c}{VIS, NIR(sec)}  \\[0.5ex] 

\hline \\[-1.8ex]

HoyosD2 5	&	23 28 41.65	&	+00 18 20.0	&	29/9/2013 	&	0.6	&	2000, 2400\\
Q2343-BM133 	&	23 46 16.18	&	+12 48 09.3	&	29/9/2013 	&	0.6	&      4000, 4640\\
Q2343-BX660 	&	23 46 29.43	&	+12 49 45.5	&	29/9/2013 	&	0.6	&	8000, 9280\\
HoyosD2 1	&	23 29 08.20	&	+00 20 40.7	&	30/9/2013	&	0.6	&		1600, 2400\\
Q2343-BX435 	&	23 46 26.36	&	+12 47 55.1	&	30/9/2013 &	0.6	&		 4000, 4640\\
Q2343-BX418 	&	23 46 18.57	&	+12 47 47.4	&	30/9/2013 	&	0.6	&	 4000, 4640\\
Q2343-BX436 	&	23 46 09.06	&	+12 47 56.0	&	30/9/2013 	&	0.6	&	 4000, 4640\\
MatsudaC3HAE3	&	02 02 37.68 	&	+01 44 33.2	&	30/9/2013 	&	0.6	& 4000, 4640\\
HoyosD2 12	&	02 28 45.05	&	+00 41 32.8	&	30/9/2013 	&	0.6	&	3200, 4800\\
\hline
\end{tabular}
\normalsize 
\end{minipage}
\end{table*}

%% file: tab/tab01.tex
\begin{table*}
 \scriptsize 
 \centering
 \begin{minipage}{140mm}
  \caption{Observed $[\Oiii]\ \lambda 5007$ and $\ha$ central wavelength  and FWHM.}
  \label{tab:tab601}
  \begin{tabular}{@{}  l c c c c c c @{}}
\hline \hline \\[-2ex]
\multicolumn{1}{c}{(1)} &
\multicolumn{1}{c}{(2)} &
\multicolumn{1}{c}{(3)} &
\multicolumn{1}{c}{(4)} &
\multicolumn{1}{c}{(5)} &
\multicolumn{1}{c}{(6)} &
\multicolumn{1}{c}{(7)} \\ 

\multicolumn{1}{c}{\bf{Name}} &
\multicolumn{1}{c}{$\mathbf{z_{hel}}$} &
\multicolumn{1}{c}{ \textbf{$\mathbf{\lambda_{c}}$ ($\mathbf{[\Oiii]\ \lambda  5007}$)}} &
\multicolumn{1}{c}{ \textbf{FWHM($\mathbf{[\Oiii]\ \lambda  5007}$)}} &
\multicolumn{1}{c}{ \textbf{$\mathbf{\lambda_{c}}$ ($\mathbf{\ha}$)}} &
\multicolumn{1}{c}{ \textbf{FWHM ($\mathbf{\ha}$)}} &
\multicolumn{1}{c}{ \textbf{Remarks\footnote{1: Object used in the analysis, 2:  log $\sigma > $ 1.8; object not used.}}} \\

\multicolumn{1}{c}{} &
\multicolumn{1}{c}{ } &
\multicolumn{1}{c}{(nm)} & 
\multicolumn{1}{c}{($\mathrm{km\ s^{-1}}$)} & 
\multicolumn{1}{c}{(nm)} & 
\multicolumn{1}{c}{($\mathrm{km\ s^{-1}}$)}  &
\multicolumn{1}{c}{ } \\[0.5ex] \hline \\[-1.8ex]

HoyosD2 5	&	0.6364	&	819.359	$\pm$	0.002	&	96.6	$\pm$	1.3	&	1073.861	$\pm$	0.012	&	113.3	$\pm$	7.8	&	1	\\
Q2343-BM133 	&	1.4774	&	1240.440	$\pm$	0.003	&	134.1	$\pm$	1.5	&	1625.881	$\pm$	0.005	&	142.6	$\pm$	2.1	&	1	\\
Q2343-BX660 	&	2.1735	&	1589.181	$\pm$	0.006	&	150.9	$\pm$	2.5	&	2082.558	$\pm$	0.012	&	---	&	1	\\
HoyosD2 1	&	0.8510	&	926.740	$\pm$	0.015	&	118.2	$\pm$	11.4	&	---	&	---	&	1	\\
Q2343-BX435 	&	2.1119	&	1557.924	$\pm$	0.034	&	171.3	$\pm$	15.6	&	2042.258	$\pm$	0.021	&	178.6	$\pm$	7.2	&	2	\\
Q2343-BX418 	&	2.3052	&	1654.870	$\pm$	0.002	&	135.3	$\pm$	1.0	&	---	&	---	&	1	\\
Q2343-BX436 	&	2.3277	&	1666.438	$\pm$	0.025	&	180.0	$\pm$	10.8	&	---	&	---	&	2	\\
MatsudaC3HAE3	&	2.2397	&	1622.102	$\pm$	0.054	&	292.7	$\pm$	23.4	&	---	&	---	&	2	\\
HoyosD2 12	&	0.6816	&	841.988	$\pm$	0.002	&	84.4	$\pm$	1.2	&	---	&	---	&	1	\\

\hline
\end{tabular}
\normalsize 
\end{minipage}
\end{table*}

%% file: tab/tab02.tex
\begin{table*}
\scriptsize
 \centering
 \begin{minipage}{160mm}
  \caption{Luminosity  and  gas velocity dispersion of high-$z$ \hii\ Galaxies obtained from the literature and from observations.}
  \label{tab:tab602} 
  \begin{tabular}{@{} l c c c c c c c c c c @{}}
\hline \hline \\[-2ex]
\multicolumn{1}{c}{(1)} &
\multicolumn{1}{c}{(2)} &
\multicolumn{1}{c}{(3)} &
\multicolumn{1}{c}{(4)} &
\multicolumn{1}{c}{(5)} &
\multicolumn{1}{c}{(6)} &
\multicolumn{1}{c}{(7)} &
\multicolumn{1}{c}{(8)} &
\multicolumn{1}{c}{(9)} &
\multicolumn{1}{c}{(10)} \\

\multicolumn{1}{c}{\bf{Name}} &
\multicolumn{1}{c}{\textbf{RA}} & 
\multicolumn{1}{c}{ \textbf{$\mathbf{\delta}$}} & 
\multicolumn{1}{c}{$\mathbf{z_{hel}}$} &
\multicolumn{1}{c}{ \textbf{$\mathbf{\log \sigma}$}} &
\multicolumn{1}{c}{ \textbf{$\mathbf{\log L}$ ($\mathbf{\hb}$)}} &
\multicolumn{1}{c}{ \textbf{$\mathbf{f}$ ($\mathbf{\hb}$)}} &
\multicolumn{1}{c}{ \textbf{$\mathbf{A_V}$ }} &
\multicolumn{1}{c}{ \textbf{$\mathbf{W}$ ($\mathbf{\ha}$)}} &
\multicolumn{1}{c}{\textbf{Ref.}} \\

\multicolumn{1}{c}{} &
\multicolumn{2}{c}{(J2000.0)} &
\multicolumn{1}{c}{} &
\multicolumn{1}{c}{($\mathrm{km\ s^{-1}}$)} & 
\multicolumn{1}{c}{($\mathrm{erg\ s^{-1}}$)} & 
\multicolumn{1}{c}{($10^{-17}\ \mathrm{erg\ s^{-1}\ cm^{-2}}$)} &
\multicolumn{1}{c}{} &
\multicolumn{1}{c}{(\AA)} & 
\multicolumn{1}{c}{} \\[0.5ex] \hline \\[-1.8ex]

Q2343-BM133	&	23 46 16.18	&	+12 48 09.31	&	1.4774	&	1.756	$\pm$	0.017	&	$42.202	^{+0.053}	_{-0.060}$	&	13.069	$\pm$	1.293	&	0.20	$\pm$	0.14	&	2245			&	1	\\
Q2343-BX418	&	23 46 18.57	&	+12 47 47.38	&	2.3052	&	1.758	$\pm$	0.016	&	$42.041	^{+0.038}	_{-0.042}$	&	3.031	$\pm$	0.116	&	0.14	$\pm$	0.10	&	1639			&	1	\\
Q2343-BX660	&	23 46 29.43	&	+12 49 45.54	&	2.1735	&	1.808	$\pm$	0.016	&	$42.024	^{+0.039}	_{-0.043}$	&	3.363	$\pm$	0.146	&	0.04	$\pm$	0.03	&	488			&	1	\\
HoyosD2-5	&	23 28 41.65	&	+00 18 20.00	&	0.6364	&	1.597	$\pm$	0.023	&	$41.393	^{+0.151}	_{-0.233}$	&	16.171	$\pm$	6.573	&	0.28	$\pm$	0.04	&	96$^a$	$\pm$	5	&	2	\\
HoyosD2-1	&	23 29 08.20	&	+00 20 40.70	&	0.8510	&	1.695	$\pm$	0.049	&	$41.692	^{+0.151}	_{-0.233}$	&	15.803	$\pm$	6.424	&	0.28	$\pm$	0.04	&	98$^a$	$\pm$	5	&	2	\\
HoyosD2-12	&	02 28 45.05	&	+00 41 32.80	&	0.6816	&	1.527	$\pm$	0.027	&	$41.297	^{+0.149}	_{-0.230}$	&	10.965	$\pm$	4.410	&	0.00	$\pm$	0.04	&	110$^a$	$\pm$	15	&	2	\\

\hline \\[-1.8ex]

HDF-BX1277	&	12 37 18.59	&	+62 09 55.54	&	2.2713	&	1.799	$\pm$	0.062	&	$41.907	^{+0.049}	_{-0.056}$	&	2.305	$\pm$	0.201	&	0.29	$\pm$	0.09	&	---	&	1	\\
Q0201-B13	&	02 03 49.25	&	+11 36 10.58	&	2.1663	&	1.792	$\pm$	0.070	&	$41.421	^{+0.039}	_{-0.043}$	&	0.845	$\pm$	0.035	&	0.01	$\pm$	0.00	&	---	&	1	\\
Q1623-BX215	&	16 25 33.80	&	+26 53 50.66	&	2.1814	&	1.845	$\pm$	0.093	&	$41.860	^{+0.061}	_{-0.072}$	&	2.283	$\pm$	0.290	&	0.41	$\pm$	0.12	&	---	&	1	\\
Q1623-BX453	&	16 25 50.84	&	+26 49 31.40	&	2.1816	&	1.785	$\pm$	0.028	&	$42.459	^{+0.094}	_{-0.120}$	&	9.076	$\pm$	2.064	&	0.84	$\pm$	0.25	&	187			&	1	\\
Q2346-BX120	&	23 48 26.30	&	+00 20 33.16	&	2.2664	&	1.792	$\pm$	0.084	&	$41.815	^{+0.042}	_{-0.046}$	&	1.875	$\pm$	0.106	&	0.02	$\pm$	0.00	&	---	&	1	\\
Q2346-BX405	&	23 48 21.22	&	+00 24 45.46	&	2.0300	&	1.699	$\pm$	0.035	&	$42.125	^{+0.035}	_{-0.039}$	&	5.009	$\pm$	0.083	&	0.03	$\pm$	0.01	&	358			&	1	\\
COSMOS-17839	&	10 00 40.96	&	+02 21 38.88	&	1.4120	&	1.664	$\pm$	0.084	&	$41.205	^{+0.298}	_{-1.848}$	&	1.472	$\pm$	1.446	&	0.33	$\pm$	0.09	&	325	$\pm$	230	&	3	\\
WISP159-134	&	20 56 30.91	&	-04 47 56.30	&	1.3000	&	1.686	$\pm$	0.045	&	$41.684	^{+0.053}	_{-0.060}$	&	5.443	$\pm$	0.532	&	0.12	$\pm$	0.09	&	314	$\pm$	36	&	4	\\
WISP173-205	&	01 55 23.64	&	-09 03 10.20	&	1.4440	&	1.834	$\pm$	0.045	&	$41.684	^{+0.062}	_{-0.072}$	&	4.196	$\pm$	0.535	&	0.09	$\pm$	0.15	&	603	$\pm$	42	&	4	\\
WISP46-75	&	22 37 56.48	&	-18 42 46.10	&	1.5040	&	1.839	$\pm$	0.066	&	$41.832	^{+0.129}	_{-0.185}$	&	5.334	$\pm$	1.794	&	0.15	$\pm$	0.43	&	245	$\pm$	28	&	4	\\
WISP22-216	&	08 52 46.29	&	+03 08 45.90	&	1.5430	&	1.641	$\pm$	0.040	&	$41.657	^{+0.054}	_{-0.062}$	&	3.350	$\pm$	0.347	&	0.00	$\pm$	0.12	&	---	&	4	\\
WISP64-2056	&	14 37 30.20	&	-01 50 51.40	&	1.6100	&	1.746	$\pm$	0.039	&	$41.716	^{+0.050}	_{-0.056}$	&	3.457	$\pm$	0.304	&	0.33	$\pm$	0.09	&	---	&	4	\\
WISP138-173	&	15 45 31.03	&	+09 33 30.00	&	2.1580	&	1.814	$\pm$	0.040	&	$42.118	^{+0.059}	_{-0.068}$	&	4.245	$\pm$	0.504	&	0.33	$\pm$	0.09	&	---	&	4	\\
WISP64-210	&	14 37 28.34	&	-01 49 54.40	&	2.1770	&	1.830	$\pm$	0.039	&	$42.043	^{+0.052}	_{-0.059}$	&	3.498	$\pm$	0.331	&	0.33	$\pm$	0.09	&	---	&	4	\\
WISP204-133	&	11 19 46.37	&	+04 10 30.80	&	2.1910	&	1.765	$\pm$	0.063	&	$41.607	^{+0.053}	_{-0.060}$	&	1.262	$\pm$	0.126	&	0.00	$\pm$	0.00	&	---	&	4	\\
WISP70-253	&	04 02 02.50	&	-05 37 19.50	&	2.2150	&	1.628	$\pm$	0.041	&	$41.590	^{+0.038}	_{-0.042}$	&	1.182	$\pm$	0.045	&	0.00	$\pm$	0.00	&	---	&	4	\\
WISP96-158	&	02 09 26.37	&	-04 43 29.00	&	2.2340	&	1.702	$\pm$	0.043	&	$41.964	^{+0.052}	_{-0.059}$	&	2.742	$\pm$	0.259	&	0.33	$\pm$	0.09	&	---	&	4	\\
WISP138-160	&	15 45 36.29	&	+09 34 26.70	&	2.2640	&	1.838	$\pm$	0.044	&	$42.318	^{+0.049}	_{-0.055}$	&	5.987	$\pm$	0.512	&	0.33	$\pm$	0.09	&	---	&	4	\\
WISP206-261	&	10 34 17.56	&	-28 30 49.80	&	2.3150	&	1.693	$\pm$	0.044	&	$42.153	^{+0.147}	_{-0.223}$	&	3.880	$\pm$	1.525	&	0.91	$\pm$	0.49	&	---	&	4	\\

\hline
\multicolumn{8}{l}{NOTE. - Units of right ascension are hours, minutes, and seconds, and units of declination are degrees,}\\
\multicolumn{8}{l}{arcminutes, and arcseconds.}\\
\multicolumn{8}{l}{References: (1)\citet{Erb2006}, (2)\citet{Hoyos2005}, (3)\citet{Maseda2014} \& (4)\citet{Masters2014}.}\\
\multicolumn{8}{l}{$^a$ $\mathbf{W}$ ($\mathbf{\hb}$).}

\end{tabular}
\normalsize 
\end{minipage}
\end{table*}

%% file: h2galcosmoV2-01.bbl
\begin{thebibliography}{39}
\expandafter\ifx\csname natexlab\endcsname\relax\def\natexlab#1{#1}\fi

\bibitem[{{Amanullah} {et~al}\mbox{.}(2010){Amanullah}, {Lidman}, {Rubin},
  {Aldering}, {Astier}, {Barbary}, {Burns}, {Conley}, {Dawson}, {Deustua},
  {Doi}, {Fabbro}, {Faccioli}, {Fakhouri}, {Folatelli}, {Fruchter}, {Furusawa},
  \& {Garavini}}]{Amanullah2010}
{Amanullah} R. {et~al.}, 2010, \apj, 716, 712

\bibitem[{{Bordalo} \& {Telles}(2011)}]{Bordalo2011}
{Bordalo} V., {Telles} E., 2011, \apj, 735, 52

\bibitem[{{Bosch}, {Terlevich} \& {Terlevich}(2002){Bosch}, {Terlevich}, \&
  {Terlevich}}]{Bosch2002}
{Bosch} G., {Terlevich} E., {Terlevich} R., 2002, \mnras, 329, 481

\bibitem[{{Calzetti} {et~al}\mbox{.}(2000){Calzetti}, {Armus}, {Bohlin},
  {Kinney}, {Koornneef}, \& {Storchi-Bergmann}}]{Calzetti2000}
{Calzetti} D., {Armus} L., {Bohlin} R.~C., {Kinney} A.~L., {Koornneef} J.,
  {Storchi-Bergmann} T., 2000, \apj, 533, 682

\bibitem[{{Ch{\'a}vez} {et~al}\mbox{.}(2012){Ch{\'a}vez}, {Terlevich},
  {Terlevich}, {Plionis}, {Bresolin}, {Basilakos}, \& {Melnick}}]{Chavez2012}
{Ch{\'a}vez} R., {Terlevich} E., {Terlevich} R., {Plionis} M., {Bresolin} F.,
  {Basilakos} S., {Melnick} J., 2012, \mnras, 425, L56

\bibitem[{{Ch{\'a}vez} {et~al}\mbox{.}(2014){Ch{\'a}vez}, {Terlevich},
  {Terlevich}, {Bresolin}, {Melnick}, {Plionis}, \& {Basilakos}}]{Chavez2014}
{Ch{\'a}vez} R., {Terlevich} R., {Terlevich} E., {Bresolin} F., {Melnick} J.,
  {Plionis} M., {Basilakos} S., 2014, \mnras, 442, 3565

\bibitem[{{Chevallier} \& {Polarski}(2001)}]{Chevallier2001}
{Chevallier} M., {Polarski} D., 2001, International Journal of Modern Physics
  D, 10, 213

\bibitem[{{Dottori}(1981)}]{Dottori1981}
{Dottori} H.~A., 1981, \apss, 80, 267

\bibitem[{{Dottori} \& {Bica}(1981)}]{Dottori1981b}
{Dottori} H.~A., {Bica} E.~L.~D., 1981, \aap, 102, 245

\bibitem[{{Erb} {et~al}\mbox{.}(2006{\natexlab{a}}){Erb}, {Steidel}, {Shapley},
  {Pettini}, {Reddy}, \& {Adelberger}}]{Erb2006b}
{Erb} D.~K., {Steidel} C.~C., {Shapley} A.~E., {Pettini} M., {Reddy} N.~A.,
  {Adelberger} K.~L., 2006{\natexlab{a}}, \apj, 647, 128

\bibitem[{{Erb} {et~al}\mbox{.}(2006{\natexlab{b}}){Erb}, {Steidel}, {Shapley},
  {Pettini}, {Reddy}, \& {Adelberger}}]{Erb2006}
{Erb} D.~K., {Steidel} C.~C., {Shapley} A.~E., {Pettini} M., {Reddy} N.~A.,
  {Adelberger} K.~L., 2006{\natexlab{b}}, \apj, 646, 107

\bibitem[{{Frieman} {et~al}\mbox{.}(2003){Frieman}, {Huterer}, {Linder}, \&
  {Turner}}]{Frieman2003}
{Frieman} J.~A., {Huterer} D., {Linder} E.~V., {Turner} M.~S., 2003, \prd, 67,
  083505

\bibitem[{{Fuentes-Masip} {et~al}\mbox{.}(2000){Fuentes-Masip},
  {Mu{\~n}oz-Tu{\~n}{\'o}n}, {Casta{\~n}eda}, \&
  {Tenorio-Tagle}}]{Fuentes-Masip2000}
{Fuentes-Masip} O., {Mu{\~n}oz-Tu{\~n}{\'o}n} C., {Casta{\~n}eda} H.~O.,
  {Tenorio-Tagle} G., 2000, \aj, 120, 752

\bibitem[{{Hicken} {et~al}\mbox{.}(2009){Hicken}, {Wood-Vasey}, {Blondin},
  {Challis}, {Jha}, {Kelly}, {Rest}, \& {Kirshner}}]{Hicken2009}
{Hicken} M., {Wood-Vasey} W.~M., {Blondin} S., {Challis} P., {Jha} S., {Kelly}
  P.~L., {Rest} A., {Kirshner} R.~P., 2009, \apj, 700, 1097

\bibitem[{{Hoyos} {et~al}\mbox{.}(2005){Hoyos}, {Koo}, {Phillips}, {Willmer},
  \& {Guhathakurta}}]{Hoyos2005}
{Hoyos} C., {Koo} D.~C., {Phillips} A.~C., {Willmer} C.~N.~A., {Guhathakurta}
  P., 2005, \apjl, 635, L21

\bibitem[{{Komatsu} {et~al}\mbox{.}(2011){Komatsu}, {Smith}, {Dunkley},
  {Bennett}, {Gold}, {Hinshaw}, {Jarosik}, {Larson}, {Nolta}, {Page},
  {Spergel}, \& {Halpern}}]{Komatsu2011}
{Komatsu} E. {et~al.}, 2011, \apjs, 192, 18

\bibitem[{{Linder}(2003)}]{Linder2003}
{Linder} E.~V., 2003, Physical Review Letters, 90, 091301

\bibitem[{{Maseda} {et~al}\mbox{.}(2014){Maseda}, {van der Wel}, {Rix}, {da
  Cunha}, {Pacifici}, {Momcheva}, {Brammer}, {Meidt}, {Franx}, {van Dokkum},
  {Fumagalli}, {Bell}, {Ferguson}, {F{\"o}rster-Schreiber}, {Koekemoer}, {Koo},
  {Lundgren}, {Marchesini}, {Nelson}, {Patel}, {Skelton}, {Straughn}, {Trump},
  \& {Whitaker}}]{Maseda2014}
{Maseda} M.~V. {et~al.}, 2014, \apj, 791, 17

\bibitem[{{Masters} {et~al}\mbox{.}(2014){Masters}, {McCarthy}, {Siana},
  {Malkan}, {Mobasher}, {Atek}, {Henry}, {Martin}, {Rafelski}, {Hathi},
  {Scarlata}, {Ross}, {Bunker}, {Blanc}, {Bedregal}, {Dom{\'{\i}}nguez},
  {Colbert}, {Teplitz}, \& {Dressler}}]{Masters2014}
{Masters} D. {et~al.}, 2014, \apj, 785, 153

\bibitem[{{Matsuda} {et~al}\mbox{.}(2011){Matsuda}, {Smail}, {Geach}, {Best},
  {Sobral}, {Tanaka}, {Nakata}, {Ohta}, {Kurk}, {Iwata}, {Bielby}, {Wardlow},
  {Bower}, {Ivison}, {Kodama}, {Yamada}, {Mawatari}, \& {Casali}}]{Matsuda2011}
{Matsuda} Y. {et~al.}, 2011, \mnras, 416, 2041

\bibitem[{{Melnick}, {Tenorio-Tagle} \& {Terlevich}(1999){Melnick},
  {Tenorio-Tagle}, \& {Terlevich}}]{Melnick1999}
{Melnick} J., {Tenorio-Tagle} G., {Terlevich} R., 1999, \mnras, 302, 677

\bibitem[{{Melnick}, {Terlevich} \& {Moles}(1988){Melnick}, {Terlevich}, \&
  {Moles}}]{Melnick1988}
{Melnick} J., {Terlevich} R., {Moles} M., 1988, \mnras, 235, 297

\bibitem[{{Melnick}, {Terlevich} \& {Terlevich}(2000){Melnick}, {Terlevich}, \&
  {Terlevich}}]{Melnick2000}
{Melnick} J., {Terlevich} R., {Terlevich} E., 2000, \mnras, 311, 629

\bibitem[{{Percival} {et~al}\mbox{.}(2010){Percival}, {Reid}, {Eisenstein},
  {Bahcall}, {Budavari}, {Frieman}, {Fukugita}, {Gunn}, {Ivezi{\'c}}, {Knapp},
  {Kron}, {Loveday}, {Lupton}, {McKay}, {Meiksin}, {Nichol}, {Pope},
  {Schlegel}, {Schneider}, {Spergel}, {Stoughton}, {Strauss}, {Szalay},
  {Tegmark}, {Vogeley}, {Weinberg}, {York}, \& {Zehavi}}]{Percival2010}
{Percival} W.~J. {et~al.}, 2010, \mnras, 401, 2148

\bibitem[{{Perlmutter} {et~al}\mbox{.}(1999){Perlmutter}, {Aldering},
  {Goldhaber}, {Knop}, {Nugent}, {Castro}, {Deustua}, {Fabbro}, {Goobar},
  {Groom}, {Hook}, {Kim}, {Kim}, {Lee}, {Nunes}, {Pain}, {Pennypacker}, \&
  {Quimby}}]{Perlmutter1999}
{Perlmutter} S. {et~al.}, 1999, \apj, 517, 565

\bibitem[{{Plionis} {et~al}\mbox{.}(2011){Plionis}, {Terlevich}, {Basilakos},
  {Bresolin}, {Terlevich}, {Melnick}, \& {Chavez}}]{Plionis2011}
{Plionis} M., {Terlevich} R., {Basilakos} S., {Bresolin} F., {Terlevich} E.,
  {Melnick} J., {Chavez} R., 2011, \mnras, 416, 2981

\bibitem[{{Riess} {et~al}\mbox{.}(1998){Riess}, {Filippenko}, {Challis},
  {Clocchiatti}, {Diercks}, {Garnavich}, {Gilliland}, {Hogan}, {Jha},
  {Kirshner}, {Leibundgut}, {Phillips}, \& {Reiss}}]{Riess1998}
{Riess} A.~G. {et~al.}, 1998, \aj, 116, 1009

\bibitem[{{Riess} {et~al}\mbox{.}(2011){Riess}, {Macri}, {Casertano},
  {Lampeit}, {Ferguson}, {Filippenko}, {Jha}, {Li}, {Chornock}, \&
  {Silverman}}]{Riess2011}
{Riess} A.~G. {et~al.}, 2011, \apj, 730, 119

\bibitem[{{Sargent} \& {Searle}(1970)}]{Sargent1970}
{Sargent} W.~L.~W., {Searle} L., 1970, \apjl, 162, L155

\bibitem[{{Siegel} {et~al}\mbox{.}(2005){Siegel}, {Guzm{\'a}n}, {Gallego},
  {Ordu{\~n}a L{\'o}pez}, \& {Rodr{\'{\i}}guez Hidalgo}}]{Siegel2005}
{Siegel} E.~R., {Guzm{\'a}n} R., {Gallego} J.~P., {Ordu{\~n}a L{\'o}pez} M.,
  {Rodr{\'{\i}}guez Hidalgo} P., 2005, \mnras, 356, 1117

\bibitem[{{Suyu} {et~al}\mbox{.}(2012){Suyu}, {Treu}, {Blandford}, {Freedman},
  {Hilbert}, {Blake}, {Braatz}, {Courbin}, {Dunkley}, {Greenhill}, {Humphreys},
  \& {Jha}}]{Suyu2012}
{Suyu} S.~H. {et~al.}, 2012, ArXiv: astro-ph/1202.4459

\bibitem[{{Suzuki} {et~al}\mbox{.}(2012){Suzuki}, {Rubin}, {Lidman},
  {Aldering}, {Amanullah}, {Barbary}, {Barrientos}, {Botyanszki}, {Brodwin},
  {Connolly}, {Dawson}, {Dey}, {Doi}, {Donahue}, {Deustua}, {Eisenhardt},
  {Ellingson}, {Faccioli}, {Fadeyev}, {Fakhouri}, {Fruchter}, {Gilbank},
  {Gladders}, {Goldhaber}, {Gonzalez}, {Goobar}, {Gude}, {Hattori}, {Hoekstra},
  {Hsiao}, {Huang}, {Ihara}, {Jee}, {Johnston}, {Kashikawa}, {Koester},
  {Konishi}, {Kowalski}, {Linder}, {Lubin}, {Melbourne}, {Meyers}, {Morokuma},
  {Munshi}, {Mullis}, {Oda}, {Panagia}, {Perlmutter}, {Postman}, {Pritchard},
  {Rhodes}, {Ripoche}, {Rosati}, {Schlegel}, {Spadafora}, {Stanford},
  {Stanishev}, {Stern}, {Strovink}, {Takanashi}, {Tokita}, {Wagner}, {Wang},
  {Yasuda}, {Yee}, \& {Supernova Cosmology Project}}]{Suzuki2012}
{Suzuki} N. {et~al.}, 2012, \apj, 746, 85

\bibitem[{{Telles}(2003)}]{Telles2003}
{Telles} E., 2003, in Astronomical Society of the Pacific Conference Series,
  Vol. 297, Star Formation Through Time, {E.~Perez, R.~M.~Gonzalez Delgado, \&
  G.~Tenorio-Tagle}, ed., p. 143

\bibitem[{{Tenorio-Tagle}, {Munoz-Tunon} \& {Cox}(1993){Tenorio-Tagle},
  {Munoz-Tunon}, \& {Cox}}]{GTT1993}
{Tenorio-Tagle} G., {Munoz-Tunon} C., {Cox} D.~P., 1993, \apj, 418, 767

\bibitem[{{Terlevich}(1997)}]{Terlevich1997}
{Terlevich} R., 1997, in Revista Mexicana de Astronomia y Astrofisica, vol. 27,
  Vol.~6, Revista Mexicana de Astronomia y Astrofisica Conference Series,
  {Franco} J., {Terlevich} R., {Serrano} A., eds., p.~1

\bibitem[{{Terlevich} \& {Melnick}(1981)}]{Terlevich1981}
{Terlevich} R., {Melnick} J., 1981, \mnras, 195, 839

\bibitem[{{Vernet} {et~al}\mbox{.}(2011){Vernet}, {Dekker}, {D'Odorico},
  {Kaper}, {Kjaergaard}, {Hammer}, {Randich}, {Zerbi}, {Groot}, {Hjorth},
  {Guinouard}, {Navarro}, {Adolfse}, {Albers}, {Amans}, {Andersen}, {Andersen},
  {Binetruy}, {Bristow}, {Castillo}, {Chemla}, {Christensen}, {Conconi},
  {Conzelmann}, {Dam}, {de Caprio}, {de Ugarte Postigo}, {Delabre}, {di
  Marcantonio}, {Downing}, {Elswijk}, {Finger}, {Fischer}, {Flores}, {Fran{\c
  c}ois}, {Goldoni}, {Guglielmi}, {Haigron}, {Hanenburg}, {Hendriks},
  {Horrobin}, {Horville}, {Jessen}, {Kerber}, {Kern}, {Kiekebusch}, {Kleszcz},
  {Klougart}, {Kragt}, {Larsen}, {Lizon}, {Lucuix}, {Mainieri}, {Manuputy},
  {Martayan}, {Mason}, {Mazzoleni}, {Michaelsen}, {Modigliani}, {Moehler},
  {M{\o}ller}, {Norup S{\o}rensen}, {N{\o}rregaard}, {P{\'e}roux}, {Patat},
  {Pena}, {Pragt}, {Reinero}, {Rigal}, {Riva}, {Roelfsema}, {Royer}, {Sacco},
  {Santin}, {Schoenmaker}, {Spano}, {Sweers}, {Ter Horst}, {Tintori}, {Tromp},
  {van Dael}, {van der Vliet}, {Venema}, {Vidali}, {Vinther}, {Vola},
  {Winters}, {Wistisen}, {Wulterkens}, \& {Zacchei}}]{Vernet2011}
{Vernet} J. {et~al.}, 2011, \aap, 536, A105

\bibitem[{{Weinberg}(2008)}]{Weinberg2008}
{Weinberg} S., 2008, {Cosmology}. Oxford University Press

\bibitem[{{Zaragoza-Cardiel} {et~al}\mbox{.}(2015){Zaragoza-Cardiel},
  {Beckman}, {Font}, {Garc{\'{\i}}a-Lorenzo}, {Camps-Fari{\~n}a}, {Fathi},
  {James}, {Erroz-Ferrer}, {Barrera-Ballesteros}, \&
  {Cisternas}}]{Zaragoza-Cardiel2015}
{Zaragoza-Cardiel} J. {et~al.}, 2015, ArXiv e-prints, 1505.01497

\end{thebibliography}
